\documentclass[twocolumn,pra,floatfix]{revtex4-1}
\usepackage{hyperref}
\usepackage{amsmath}
\usepackage{amsfonts}
\usepackage{amsbsy}
\usepackage{xcolor}
\usepackage{soul}
\usepackage{graphicx}
\usepackage{epsfig}

\newcommand{\bra}[1]{\ensuremath{\left\langle #1 \right\vert}}
\newcommand{\ket}[1]{\ensuremath{\left\vert #1 \right\rangle}}
\newcommand{\quantmean}[1]{\ensuremath{\left\langle #1 \right\rangle}}

\newcommand{\boldvec}[1]{\ensuremath{\boldsymbol{\mathbf{#1}}}}
\newcommand{\spvec}[1]{\ensuremath{\boldvec{#1}}}
\newcommand{\unitvec}[1]{\ensuremath{\hat{\boldvec{#1}}}}

\newcommand{\pol}{\unitvec{e}}
\newcommand{\rv}{{\spvec{r}}}

\newcommand{\Rv}{\spvec{R}}
\newcommand{\Ev}{{\spvec{E}}}
\newcommand{\Evhat}{\hat{{\spvec{E}}}}
\newcommand{\Dv}{{\spvec{D}}}
\newcommand{\Dvhat}{\hat{{\spvec{D}}}}
\newcommand{\Pv}{{\spvec{P}}}
\newcommand{\Pvhat}{\hat{{\spvec{P}}}}

\newcommand{\dv}{\spvec{d}}

\newcommand{\Dc}{\mathcal{D}}
\newcommand{\Hc}{\mathcal{H}}

\newcommand{\qv}{{\bf q}}

\newcommand{\eo}{\epsilon_0}
\newcommand{\radKernel}{\ensuremath{\mathsf{G}}}

\newcommand{\beq}{\begin{equation}}
\newcommand{\eeq}{\end{equation}}
\newcommand{\bea}{\begin{eqnarray}}
\newcommand{\eea}{\end{eqnarray}}
\newcommand{\<}{\langle}
\renewcommand{\>}{\rangle}

\newcommand{\Pc}{{\cal P}}
\newcommand{\Pcv}{\boldsymbol{{\cal P}}}

\newcommand{\commentout}[1]{{}}

\newcommand{\br}{{\bf r}}
\newcommand{\bE}{{\bf E}}

\newcommand{\cpsi}{\ensuremath{\hat{\psi}^\dagger}}
\newcommand{\dpsi}{\ensuremath{\hat{\psi}}}
\newcommand{\av}[1]{\ensuremath{\langle #1 \rangle}}

\newcommand{\stochx}{\ensuremath{X}}
\newcommand{\stochxv}{\spvec{\stochx}}

\begin{document}
\author{Mark D. Lee}
\affiliation{Mathematical Sciences, University of Southampton, Southampton SO17 1BJ, United Kingdom}
\author{Stewart D. Jenkins}
\affiliation{Mathematical Sciences, University of Southampton, Southampton SO17 1BJ, United Kingdom}
\author{Janne Ruostekoski}
\affiliation{Mathematical Sciences, University of Southampton, Southampton SO17 1BJ, United Kingdom}
\date{\today }
\title{Stochastic methods for light propagation and recurrent scattering in saturated and nonsaturated atomic ensembles}
\begin{abstract}

We derive equations for the strongly coupled system of light and dense atomic
ensembles. The formalism includes an arbitrary internal level
structure for the atoms and is not restricted to weak excitation of atoms by
light. In the low light intensity limit for atoms with a single electronic ground
state, the full quantum field-theoretical representation of the model can be solved exactly by means of classical stochastic electrodynamics simulations for stationary atoms that represent cold atomic
ensembles.
Simulations for the optical response of
atoms in a quantum degenerate regime require one to
synthesize a stochastic ensemble of atomic positions that generates the
corresponding quantum statistical position correlations between the atoms.
In the case of multiple ground levels or at light intensities where
saturation becomes important, the classical simulations require approximations that neglect
quantum fluctuations between the levels. We show how the model is extended to incorporate corrections due to quantum fluctuations that result from virtual scattering processes.
In the low light intensity
limit we illustrate the simulations in a system of atoms in a Mott-insulator state in
a 2D optical lattice, where recurrent scattering of light induces strong interatomic
correlations.  These correlations result in collective many-atom sub- and
super-radiant states and a strong dependence of the response on the spatial
confinement within the lattice sites.

\end{abstract}


\maketitle

\section{Introduction}

Interactions mediated by electromagnetic fields can induce correlations
between resonant emitters in systems that would not otherwise show any correlation
effects. This behavior generally occurs when the resonant dipole-dipole interactions
between the emitters are sufficiently strong -- when the
spacing between emitters in homogeneously broadened samples
is less than a wavelength~\cite{JavanainenMFT}.

Such cooperative
phenomena~\cite{Lehmberg1970a,Lehmberg1970b,Ishimaru1978,vantiggelen90,Morice1995a,Ruostekoski1997a,
Ruostekoski1997b,Rusek96,Javanainen1999a,Ruostekoski1999a,BerhaneKennedyfer,Clemens2003a,pinheiro04,
dalibardexp,JenkinsLongPRB,Jenkins2012a,Olmos13,Castin13,evers13,Javanainen2014a,Pellegrino2014a,JavanainenMFT,
Skipetrov14,Bettles_lattice,Ritsch_lattice} cannot
be described by a model of dipoles which scatter independently. They are intimately
linked to the presence of recurrent (or dependent) scattering; when a photon scatters more than
once off the same emitter. The resulting interactions produce collective excitation
eigenmodes, each with a distinct resonance frequency and decay
rate. Some of these modes are superradiant, in which case the emission rates compared with those
of a single isolated atom are enhanced by collective interactions. In other modes, the radiative emission is
subradiant with the collective emission rates that are weaker than that of an isolated atom.
In a strongly-coupled response of large ensembles of resonant emitters
the ratio between the most sub- and super-radiant decay rates can exceed several
orders of magnitude~\cite{Rusek96,JenkinsLongPRB} if the average separation between the emitters is
sufficiently small. As well as the magnitude of the
atom density, the response can be sensitive to details such as atom
statistics~\cite{Morice1995a,Ruostekoski1999a,bcs1}, the degree of inhomogeneous
broadening~\cite{Javanainen2014a,JenkinsRuostekoskiPRB2012b}, or spatial disorder
that can lead to light localization~\cite{Segev2013}. Light-induced correlations between the atoms due to resonant dipole-dipole interactions
can be entirely classical and may result in a failure~\cite{Javanainen2014a} of
standard textbook theories~\cite{Jackson,BOR99} of the electrodynamics
of a polarizable medium that are based on effective continuous medium models.

Spectroscopic experiments can now be performed with optically
dense atomic ensembles at increasing atom
densities~\cite{BalikEtAl2013,KEM14,Pellegrino2014a,Keaveney2012,BIE10,ChabeEtAlPRA2014,wilkowski}.
It has been predicted that correlation phenomena due to light-mediated
interactions can emerge in cold atomic ensembles already at surprisingly low
atom densities~\cite{JavanainenMFT}. In particular, in cold alkali-metal atomic
gases with densities of over $10^{14}$cm$^{-3}$ the observation of correlated
scattering should be achievable~\cite{BalikEtAl2013,Pellegrino2014a}. Strong
dipole-dipole interactions in cold ensembles of highly-excited Rydberg
atoms~\cite{TongEtAlPRL2004,HeidemannEtAlPRL2007,PritchardEtALPRL2010,WilkEtAlPRL2010,SchemppEtAlPRL2010}
also provide a possible platform to explore cooperative coupling between atoms.
Strong dipole-dipole interactions may need to be considered to an increasing
extent in quantum technologies. For example, collective phenomena, such as
superradiant emission, can be important in implementations of quantum memories
and quantum interfaces between light and cold atom
systems~\cite{ChaneliereEtAlPRL2006}. The advent of metamaterials, arrays of
artificially fabricated magnetodielectric resonators whose electromagnetic
response and positions are controlled by design, provide a way to exploit
cooperative interactions, for example, to focus light or image a system beyond
the diffraction limit~\cite{LemoultPRL10}, to produce narrow transmission
resonances~\cite{FedotovEtAlPRL2010,JenkinsLineWidthNJP,CAIT}, or to create a broad
area coherent light source driven by an electron beam~\cite{AdamoEtAlPRL2012}.

Previous studies of light-atom interactions have provided a full field-theoretical formulation~\cite{Ruostekoski1997a,Ruostekoski1997b} where the atoms and light
were treated in terms of quantum fields throughout the analysis.  It was shown, specifically in the limit of low light intensity, that the strong light-matter
coupling leads to a hierarchy of equations of motion for the correlation functions of atomic density and polarization. In the resulting hierarchy, atomic
polarization density is coupled to a two-atom correlation function which is then coupled to a three-atom correlation function, etc. General solutions to such
a hierarchy of equations even in the case of small atom numbers are challenging, although perturbative solutions may be derived, e.g., in inhomogeneously broadened
samples~\cite{Javanainen2014a}, or in the limit of low atom density $\rho$ when $\rho/k^3\ll 1$~\cite{Morice1995a,Ruostekoski1999a,bcs1}, where $k$
denotes the resonant wave number of light. An alternative approach, based on a stochastic Monte-Carlo simulation technique, was presented in Ref.~\cite{Javanainen1999a}
by considering a simplified 1D scalar electrodynamics of light-atom interactions. The dynamics of the hierarchy for the equations of motion in the limit of low light intensity was
reproduced to all orders by solving a system of classical electrodynamics equations for point dipoles at fixed positions -- these spatial positions are then stochastically sampled
from appropriate spatial distributions for each realization (for a classical gas the atomic positions are independently distributed and for quantum degenerate atoms they are sampled from the joint many-body probability distribution).
Such a coupled dipole model between pointlike atoms was explicitly shown to provide an exact solution for the low light intensity scalar theory with stationary atoms at arbitrary
atom densities by means of a classical electrodynamics simulation where all the recurrent scattering processes are fully accounted for.
Analogous techniques to simulate classical electrodynamics of systems comprising discrete emitters have
been applied to a variety of systems, for example, to scatterers at fixed
positions~\cite{JenkinsLongPRB,JenkinsLineWidthNJP}, and to scatterers whose
positions are stochastically distributed \cite{pinheiro04,
dalibardexp,Jenkins2012a,Pierrat2010,Castin13,Javanainen2014a,JavanainenMFT}. These
simulations considered the low light intensity limit and involved either emitters
with an isotropic polarizability or emitters whose dipole orientations were all
fixed in the same direction (e.g., two-level atoms) or were averaged over the
spatial directions.
Situations typically encountered in experiments on cold atomic ensembles, however, involve atoms with an internal hyperfine level structure and are
not limited only to low light intensities.

In this paper we derive general stochastic simulations techniques for the
light-atom interactions in cold atomic ensembles based on a quantum
field-theoretical analysis for the coupled theory of atoms and light. We
incorporate a multiple level structure of the atoms, full vector properties of
the electromagnetic fields, and saturation of the atoms due to illuminating
light. We show that the Monte-Carlo simulation technique, and the corresponding
equations of motion for the atoms in each stochastic realization, can be derived
by a methodologically simple procedure that is based on taking expectation
values of the atoms when they are conditioned on a given set of fixed spatial
positions. This allows us to show explicitly when the entire optical response of
the atomic ensemble can be represented exactly by a relatively simple set of
coupled equations for the atomic dipoles and when more complex theories are
required in order to deal with classical and quantum-mechanical light-induced
correlations between the atoms. We specifically show that in the low light
intensity limit and for atoms with only a single electronic ground state, the
full optical response can be represented exactly (within the statistical
accuracy) in terms of classical electrodynamics equations of motion for coupled
point dipoles driven by an incident field, when the positions of the dipoles are
stochastically sampled according to the statistical position correlations of the
atoms and ensemble averaged over many such realizations. In the presence of a
Zeeman level structure where multiple electronic ground states can couple to the
light, we show that the derivation of similar classical electrodynamics
equations of motion requires an approximate treatment of fluctuations due to
repeated virtual photon exchanges between atoms occupying different ground
levels. We introduce a classical approximation where the recurrent (or
dependent) scattering between the atoms is included, but nonclassical higher
order correlations between internal levels that involve fluctuations between
ground-state coherences of different levels and the polarization are factorized.
The classical approximation correctly reproduces all the atomic ground-state
population expectation values that, in the limit of low light intensity, remain
constant (unchanged from their initial values) even in the exact quantum
solution. We show how one can go beyond this level of approximation by
progressively incorporating higher order quantum fluctuation effects.

Our derivation for the stochastic techniques for the electrodynamics simulations
is not only restricted to the low light intensity limit. Once saturation becomes
important and the population of the electronically excited levels can no longer
be neglected, we show that we can adapt the stochastic simulations to provide a
tractable coupled-dipole approach that scales favorably with the atom number and
where the internal level dynamics is treated semiclassically. This is obtained
by using factorization approximations for certain correlation functions, in a
manner analogous to the factorization approximation introduced in the low light
intensity case of multiple ground levels.

In order to illustrate briefly the stochastic electrodynamics simulations in practise, we also consider a specific numerical example.  In the low light intensity limit, we calculate the optical
response and collective excitation eigenmodes of an atomic gas confined in a 2D optical lattice by extending the model introduced in Ref.~\cite{Jenkins2012a}, where the control of atomic
positions was used to engineer highly localized subwavelength-scale optical excitations in the atomic system. We show that the system supports collective excitation eigenmodes that exhibit a broad range
of super- and sub-radiant decay rates, and we examine how this distribution depends on the lattice spacing and confinement strength. Classical electrodynamics simulations that treat multilevel $^{87}$Rb atoms for fluorescence and coherent light transmission are reported elsewhere~\cite{Pellegrino2014a,Jennewein_trans,Jenkins_thermshift}.

In Sec.~\ref{sec:general-theory-free} we present the formalism necessary to
describe the collective scattering problem for atoms in free space, and derive
general expressions for the optical response.  We then introduce the Monte-Carlo
sampling method, firstly for low light intensity in Sec.~\ref{sec:lowintensity},
before including saturation in Sec.~\ref{sec:highintensity}.  Finally, we
provide the numerical example in Sec.~\ref{sec:mott-insulator-state}.

\section{General theory}
\label{sec:general-theory-free}

In this section we derive equations for the collective optical
response of an ensemble of atoms.  We start with a quantum field-theoretical formalism introduced in
Refs.~\cite{Ruostekoski1997a,Ruostekoski1997b,bcs2}, and extend the formalism beyond the low light-intensity limit
and by systematically including multiple internal level structure.

\subsection{Hamiltonian}
\label{sec:hamiltonian_gen_the}

We begin with a nonrelativistic Hamiltonian formalism of electrodynamics in the dipole
approximation for atoms, expressing the Hamiltonian in the {\it length} gauge obtained
in the Power-Zienau-Woolley
transformation~\cite{PowerZienauPTRS1959,Woolley1971a,PowerBook,CohenT}.
The atoms and light are treated in terms of quantum fields throughout the analysis.
In second quantization formalism for the electronic ground and
excited states of the atoms we account for the sublevel structure, so that the
corresponding field operators read $\hat\psi_{g\nu}(\rv)$ and $\hat\psi_{e\eta}(\rv)$,
respectively. Here the second subscript refers to the sublevel index, such that the
complete set of atomic basis states is formed by $|g,\nu\>$ and $|e,\eta\>$.

We write the Hamiltonian density as
\beq
\Hc = \Hc_g+ \Hc_e+\Hc_F + \Hc_D + \Hc_P\,.
\eeq
The atomic internal level energy is expressed by the first two terms, for the example
of the linear Zeeman shift caused by a magnetic field of strength $B$ these are given by
\begin{align}
\Hc_g & =\hat\psi^\dagger_{g\nu}({\bf r})\big( {\mu_BB}  g_l^{(g)}\nu \big) \hat\psi_{g\nu}({\bf r})\,,\\
\Hc_e & =\hat\psi^\dagger_{e\eta}({\bf r})\big( {\mu_BB}  g_l^{(e)}\eta +\hbar\omega_0 \big) \hat\psi_{e\eta}({\bf r})\,,
\end{align}
where $g^{(g)}_l$ and $g^{(e)}_l$ are the Land\'{e}
g-factors~\cite{BransdenJoachain} for levels $g$ and $e$, and $\omega_0$ is the
resonance frequency of the $\ket{g}\leftrightarrow\ket{e}$ transition in the
absence of any magnetic field.

For the electromagnetic field we introduce the mode frequency and the photon
annihilation operator as $\omega_q$ and $\hat a_q$, respectively. The mode index
$q$ incorporates both the wave vector ${\bf q}$ and the transverse polarization
$\pol_q$. In the length gauge it is beneficial to represent the light amplitude
by the electric displacement $\Dvhat(\rv)$ which is
the basic dynamical variable for light~\cite{CohenT}.
We write it as a sum of the positive and
negative frequency components $\Dvhat=\Dvhat^++\Dvhat^-$, with $\Dvhat^+=[\Dvhat^-]^\dagger$,
and with
\beq
\Dvhat^+(\rv)= \sum_q \zeta_q \pol_q \hat a_q e^{i\qv\cdot\rv},\quad
\zeta_q=\sqrt{\frac{\hbar \epsilon_0\omega_q}{2 V}}\,,
\label{eq:Dquantisation}
\eeq
where $V$ denotes the quantization volume.
The Hamiltonian for the free electromagnetic field energy then reads
\beq
H_F=\int d^3r\, \Hc_F = \sum_q \hbar \omega_q \hat a^\dagger_q \hat a_q\,.
\eeq

The atomic polarization $\Pvhat(\rv)$ interacts with the electric displacement and we have
\beq
\Hc_D=-{1\over\eo}\,\Pvhat(\rv)\cdot\Dvhat(\rv) \,.
\label{h2}
\eeq
We also have in the Hamiltonian the polarization self-energy term:
\beq
\Hc_{\rm P}={1\over2\eo}\,\Pvhat(\rv)\cdot\Pvhat(\rv)\,.
\eeq
For nonoverlapping atomic point dipoles such a term vanishes.

In terms of our second quantized field operators, the positive frequency component of
the polarization can be written in terms of contributions from differing sublevel
transitions as
\begin{align}
\Pvhat^+(\rv) &= \sum_{\nu,\eta} \dv_{g\nu e\eta}\hat\psi^\dagger_{g\nu}(\rv)
\hat\psi_{ e\eta}(\rv) \equiv \sum_{\nu,\eta}
\Pvhat^+_{\nu\eta}(\rv)\,,
\label{pol}\\
\Pvhat^+_{\nu\eta}(\rv) &\equiv \dv_{g\nu e\eta}\hat\psi^\dagger_{g\nu}(\rv)
\hat\psi_{ e\eta}(\rv)\,,
\label{polcomp}
\end{align}
where $\dv_{g\nu e\eta}$ represents the dipole matrix element
for the transition $|e,\eta\>\rightarrow |g,\nu\>$
\beq
\dv_{g\nu e\eta}\equiv \Dc \sum_\sigma \pol_{\sigma}
\< e\eta;1g|1\sigma;g\nu\> \equiv \Dc \sum_\sigma \pol_{\sigma} {\cal C}_{\nu,\eta}^{(\sigma)} \,.
\label{dipole}
\eeq
Here the sum is over the unit circular polarization vectors $\sigma=\pm1,0$, and
${\cal C}_{\nu,\eta}^{(\sigma)}$ denote the Clebsch-Gordan coefficients of the
corresponding optical transitions. The reduced dipole matrix element is
represented by ${\cal D}$
(here chosen to be real) and $\dv_{e\eta g\nu}=\dv_{g\nu e\eta}^*$. The light
fields with the polarizations $\pol_\pm$ and $\pol_0$ drive the transitions
$|g,\nu\>\rightarrow|e,\nu\pm1\>$ and $|g,\nu\> \rightarrow|e,\nu\>$,
respectively, in such a way that only the terms $\sigma=\eta-\nu$ in Eq.~\eqref{dipole}
are nonvanishing.

\subsection{Light and matter fields}

We assume that the atoms are illuminated by a near-monochromatic incident laser at the
frequency $\Omega$. In the following we assume that all the relevant quantities are
expressed as slowly varying amplitudes by explicitly factoring out the dominant laser
frequency oscillations by writing $\hat \psi_{e\eta}\rightarrow e^{-i\Omega t}
\hat\psi_{e\eta}$, $\Dvhat^+\rightarrow e^{-i\Omega t} \Dvhat^+$, etc.

\subsubsection{Scattered light}

The light field may be obtained by integrating the equations for motion for the
photon operator $\hat a_q$, derived from the full Hamiltonian. The electric
displacement can be expressed as a sum $\Dvhat^+({\bf r})=\Dvhat^+_F({\bf r})+
\Dvhat^+_S({\bf r})$ of the free field part $\Dvhat^+_F({\bf r})$ that would
exist in the absence of the matter, and the scattered field $\Dvhat^+_S({\bf
r})$. From
\beq
\Dvhat(\rv)=\eo \Evhat(\rv)+\Pvhat(\rv)\,,
\eeq
we can write the electric field $\epsilon_0 \Evhat^+(\rv) = \Dvhat^+_F({\bf r})
+  \Dvhat^+_S({\bf r}) - \Pvhat^+(\rv)$ as the integral expression~\cite{Ruostekoski1997a} ($k=\Omega/c$)
\begin{align}
\eo\Evhat^+({\bf r})& = \Dvhat^+_F({\bf r}) +
\int d^3r'\,
{\sf G}({\bf r}-{\bf r'})\,\Pvhat^+({\bf r}')\,,
\label{eq:MonoD}\\
{\sf G}_{ij}({\bf r})& =
\left[ {\partial\over\partial r_i}{\partial\over\partial r_j} -
\delta_{ij} {\bbox \nabla}^2\right] {e^{ikr}\over4\pi r}
-\delta_{ij}\delta({\bf r})\,.
\label{eq:GDF}
\end{align}
In Eq.~(\ref{eq:MonoD}) the monochromatic dipole radiation kernel
${\sf G}(\rv)$~\cite{Jackson} provides the radiated field at $\rv$ from
a dipole with the amplitude $\hat{\bf d}$ residing at the origin:
\begin{align}
{\sf G}&({\bf r})\,\hat{\bf d}=
{k^3\over4\pi}
\left\{ (\hat{\bf n}\!\times\!\hat{\bf d}
)\!\times\!\hat{\bf n}{e^{ikr}\over kr}\right.\nonumber\\
&\left. +[3\hat{\bf n}(\hat{\bf n}\cdot\hat{\bf d})-\hat{\bf d}]
\bigl[ {1\over (kr)^3} - {i\over (kr)^2}\bigr]e^{ikr}
\right\}-{\hat{\bf d}\,\delta({\bf r})\over3}\,,
\label{eq:DOL}
\end{align}
where $\hat{\bf n} = {{\bf r}/ r}$.  When integrating over a volume including the
origin, the term in Eq.~{(\ref{eq:DOL})} proportional to $1/r^3$ does not absolutely
converge.  Following Ref.~\cite{Ruostekoski1997a} we interpret Eq.~\eqref{eq:DOL} in
such a way that the integral over an infinitesimal volume enclosing the origin
of the term inside the curly brackets vanishes.

\subsubsection{Evolution of atomic fields}

We may also derive the equations of motion for the excited and ground state atomic fields
\begin{align}
\dot{\!\hat\psi}_{e\eta}({\bf r} ) &=  i \Delta_{e\eta}\,{\hat\psi}_{e\eta}({\bf r} ) +
{i\over\hbar}\,{\bf d}_{e\eta g\nu}\cdot\Evhat^+({\bf r} )\,\hat\psi_{g\nu}({\bf r}
)\,,
\label{eq:EXF}\\
\dot{\hat\psi}_{g\nu}({\bf r} ) &=  i \Delta_{g\nu}\,{\hat\psi}_{g\nu}({\bf r})+ {i\over\hbar}\,
\Evhat^-({\bf r} )\cdot{\bf d}_{g\nu e\eta}\,{\hat\psi}_{e\eta}({\bf r})
\,.
\label{eq:GRF}
\end{align}
Here in Eq.~\eqref{eq:EXF} we implicitly sum over all $\nu$ and in Eq.~\eqref{eq:GRF} over all $\eta$. Similarly, in the remainder of the paper we
take repeated indices in expressions to indicate an implicit summation.
The atom-light detuning is denoted by
$\Delta_{e\eta}=\Omega-(\omega_0+{\mu_BB}  g_l^{(e)}\eta/\hbar)$ and
$\Delta_{g\nu}=-{\mu_BB}  g_l^{(g)}\nu/\hbar$. For reasons that become obvious later on, we would like to place all the atomic
field operators in the normal order. In order to evaluate the free field
operator expectation values, we will place $\Dvhat^+_F({\bf r})$ farthest to
right and $\Dvhat^-_F({\bf r})$ farthest to the left. Since the initial free
field is assumed to be in a coherent state, the expectation values then produce
multiplicative classical coherent fields~\cite{WallsMilburn}. Using Eq.~\eqref{eq:MonoD}, for \eqref{eq:EXF} we obtain
\begin{align}
\dot{\!\hat\psi}_{e\eta}&({\bf r}) =  i \Delta_{e\eta}\,{\hat\psi}_{e\eta}({\bf r}) +
{i\over\hbar\epsilon_0}\,{\bf d}_{e\eta g\nu}\cdot\Dvhat_F^+({\bf r})\,\hat\psi_{g\nu}({\bf r}) \nonumber\\
&+ {i\over\hbar\epsilon_0}\,{\bf d}_{e\eta g\nu}\cdot\int d^3 r' {\sf G}(\rv-\rv') \Pvhat^+(\rv') \,\hat\psi_{g\nu}({\bf r})\,.
\label{newpsie_unordered}
\end{align}
We show in App.~\ref{app:commutator} that we can evaluate the commutator
\beq
[\Dvhat_F^+({\bf r}),\hat\psi_{g\nu}({\bf r'})]= \big[{\sf G}(\rv-\rv')+\delta(\rv-\rv')\big]
{\bf d}_{g\nu e\zeta} \hat\psi_{e\zeta} ({\bf r'})\,. \label{eq:DFpsicomm}
\eeq
Taking the limit $\rv'\rightarrow\rv$ in the field-theory version of the Born and
Markov approximations~\cite{Ruostekoski1997a}, and substituting into
\eqref{newpsie_unordered} yields
\begin{align}
\dot{\!\hat\psi}_{e\eta}&({\bf r}) =  (i \Delta_{e\eta}-\gamma){\hat\psi}_{e\eta}({\bf r}) +
{i\over\hbar\epsilon_0}\,\hat\psi_{g\nu}({\bf r}) \,{\bf d}_{e\eta g\nu}\cdot\Dvhat_F^+({\bf r})\nonumber\\
&+ {i\over\hbar\epsilon_0}\,\dv_{e\eta g\nu}\cdot\int d^3 r' {\sf G}'(\rv-\rv') \Pvhat^+(\rv') \,\hat\psi_{g\nu}({\bf r})\,.
\label{newpsie}
\end{align}
The spontaneous linewidth $\gamma$ results from the imaginary part of the commutator limit and is given by
\beq
\gamma \equiv {\Dc^2 k^3\over 6\pi\hbar\eo}\,. \label{eq:spontlinewidth}
\eeq
On the other hand, the divergent real part of the commutator limit contributes
to the Lamb shift and is assumed to be included in the detuning $\Delta$.  We
have introduced the propagator
\begin{equation}
{\sf G}'_{ij}(\rv) \equiv {\sf G}_{ij}(\rv)+\frac{\delta_{ij}\delta(\rv)}{3},
\label{eq:Gprime}
\end{equation}
explicitly canceling the contact term in ${\sf G}(\rv)$, such that the integral of
${\sf G}'(\rv)$ over an infinitesimal volume about the origin vanishes.  Since the
hard-core of real interatomic potentials prevents the overlap of cold atoms, the
contact term is inconsequential for real applications.  In fact, even for ideal
point dipoles it can be shown in the low light intensity limit that the light
matter dynamics is independent of the contact term~\cite{Ruostekoski1997b}.

In the equation of motion for $\hat\psi_{g\nu}({\bf r})$ the free field is
already in the correct order and we obtain
\begin{align}
\dot{\dpsi}_{g\nu}&(\rv)=i\Delta_{g\nu}\dpsi_{g\nu}(\rv)+{i\over\hbar\epsilon_0}
\Dvhat^-_F(\rv)\cdot\dv_{g\nu e\eta}\dpsi_{e\eta}(\rv) \nonumber \\
&+{i\over\hbar\epsilon_0}\dv_{g\nu e\eta}\cdot\int d^3r'\,{\sf G}'^*({\bf r}-{\bf r'})
\Pvhat^-(\rv')\dpsi_{e\eta}(\rv).\label{eq:dpsigdt}
\end{align}

When we derive the evolution equations for the atomic polarization and the level
populations that incorporate the internal level structure of the atoms, it is
useful to introduce the following tensor
\beq
{\sf P}^{\nu\eta}_{\mu\tau} \equiv {\dv_{g\nu e\eta}
\dv_{e\mu g\tau}\over {\cal D}^2}
= \sum_{\sigma,\varsigma}
\pol_{\sigma}\pol_{\varsigma}^* {\cal C}_{\nu,\eta}^{(\sigma)}{\cal
C}_{\tau,\mu}^{(\varsigma)}
\,.
\label{pro}
\eeq
The components of ${\sf P}^{\nu\eta}_{\mu\tau}$ depend on the vector properties
of the atomic level structure weighted by the Clebsch-Gordan coefficients (here
taken to be real).

Combining Eqs.~\eqref{newpsie} and~\eqref{eq:dpsigdt}, and rewriting in normal order, the
optical response of the system can be encoded in the equations of motion for the
polarization operator component $\Pvhat^+_{\nu\eta}(\rv)$, and ground and excited
state coherence operators $\cpsi_{g\nu}\dpsi_{g\eta}$, $\cpsi_{e\nu}\dpsi_{e\eta}$
\begin{subequations}
\begin{align}
{d\over dt}\,&\Pvhat^+_{\nu\eta} =
(i\bar{\Delta}_{g\nu e\eta}-\gamma)\Pvhat^+_{\nu\eta}
+ i\xi \hat\psi^\dagger_{g\nu}\hat\psi_{g\tau} {\sf P}^{\nu\eta}_{\eta\tau}
 \Dvhat^+_F \nonumber\\
&+ i\xi
\int d^3r'\,{\sf P}^{\nu\eta}_{\eta\tau}
{\sf G}'({\bf r}-{\bf r'})\,\hat\psi^\dagger_{g\nu}\Pvhat^+(\rv')
\hat\psi_{g\tau}\nonumber\\
& -i\xi
\int d^3r'\,{\sf P}^{\nu\eta}_{\tau\nu}
{\sf G}'({\bf r}-{\bf r'})\,\hat\psi^\dagger_{e\tau}\Pvhat^+(\rv')
\hat\psi_{e\eta}\nonumber\\
&- i\xi
\hat\psi^\dagger_{e\tau}\hat\psi_{e\eta}{\sf P}^{\nu\eta}_{\tau\nu}
\Dvhat^+_F \label{Pfullnew}
\,, \\
{d\over dt}\,&\cpsi_{g\nu}\dpsi_{g\eta} =
i\bar{\Delta}_{g\nu g\eta}
\cpsi_{g\nu}\dpsi_{g\eta}
\nonumber \\
&+
2\gamma {\cal C}_{\eta,(\eta+\sigma)}^{(\sigma)}{\cal C}_{\nu,(\nu+\sigma)}^{(\sigma)}
\cpsi_{e(\nu+\sigma)}\dpsi_{e(\eta+\sigma)}\nonumber
\\
& -i\frac{\xi}{\Dc^2}\cpsi_{e\tau}\dpsi_{g\eta}\dv_{e\tau g\nu}\cdot
\Dvhat^+_F
+i\frac{\xi}{\Dc^2}\Dvhat^-_F\cdot\dv_{g\eta
e\tau}\cpsi_{g\nu}\dpsi_{e\tau} \nonumber \\
&-i\frac{\xi}{\Dc^2}\int d^3r'\,\dv_{e\tau g\nu}\cdot{\sf G}'({\bf r}-{\bf r'})
\,\cpsi_{e\tau}\Pvhat^+(\rv')\dpsi_{g\eta} \nonumber\\
&+i\frac{\xi}{\Dc^2}\int d^3r'\,\dv_{g\eta e\tau}\cdot{\sf G}'^*({\bf r}-{\bf r'})
\,\cpsi_{g\nu}\Pvhat^-(\rv')\dpsi_{e\tau}\,, \label{eq:drhog_full}\\
{d\over dt}\,&\cpsi_{e\nu}\dpsi_{e\eta} =
(i\bar{\Delta}_{e\nu e\eta}
-2\gamma)\cpsi_{e\nu}\dpsi_{e\eta}
\nonumber \\
&+i\frac{\xi}{\Dc^2}\cpsi_{e\nu}\dpsi_{g\tau}\dv_{e\eta g\tau}\cdot
\Dvhat^+_F-i\frac{\xi}{\Dc^2}\Dvhat^-_F\cdot
\dv_{g\tau e\nu}\cpsi_{g\tau}\dpsi_{e\eta} \nonumber \\
&-i\frac{\xi}{\Dc^2}\int d^3r'\,
\dv_{g\tau e\nu}\cdot{\sf G}'^*({\bf r}-{\bf r'})
\,\cpsi_{g\tau}\Pvhat^-(\rv')\dpsi_{e\eta} \nonumber \\
&+i\frac{\xi}{\Dc^2}\int d^3r'\, \dv_{e\eta g\tau}\cdot
{\sf G}'({\bf r}-{\bf r'})\,\cpsi_{e\nu}\Pvhat^+(\rv')\dpsi_{g\tau}\,
\label{eq:drhoe_full}.
\end{align}
\label{eq:fullresponseequations}
\end{subequations}
Here the repeated indices $\tau,\sigma$ indicate
implicit summations over the corresponding sublevels and polarizations, and we define
$\bar{\Delta}_{a\nu b\eta}= \Delta_{b\eta}-\Delta_{a\nu}$ ($a,b=g,e$). In
Eqs.~{(\ref{eq:fullresponseequations})} we have introduced a notational convention where we show
explicitly only the nonlocal
position dependence $\rv'$, and the local coordinate $\rv$ of the fields is suppressed.
We follow this convention in the rest of the paper whenever
the equations would otherwise become notably longer.
We have also defined
\beq
\xi\equiv \frac{\Dc^2}{\hbar\eo}\,.
\eeq

The difficulty in solving the optical response analytically is now evident. Taking
expectation values of both sides of Eqs.~{(\ref{eq:fullresponseequations})}, we
see that the one-body correlation functions of interest depend on integrals over
two-body correlation functions. In turn, the two-body correlation functions depend
on integrals over three-body correlations, and so it continues up to the $N$th
order correlation functions.  We give explicit expressions in
App.~\ref{sec:monte-carlo-eval} for the resulting hierarchy of equations of motion
for correlation functions in the limit of low light intensity. Outside of this
limit, when all correlation functions must be included, the complexity of the
coupled equations grows even more rapidly as a function of the number of atoms and
internal levels. In practice, even for systems of a few atoms in the low light
intensity limit one requires some form of truncation  in order to solve the
hierarchy of equations. Such approximations to the light-induced correlations may
be obtained as a perturbative expansion in the small parameter $\rho/k^3$, which
is valid at low atom
densities~\cite{Morice1995a,Ruostekoski1999a,vantiggelen90,bcs1}, or the Doppler
broadening of the atoms in inhomogeneously broadened
samples~\cite{Javanainen2014a}.  In contrast, the stochastic techniques which we
describe in the following sections provide a computationally efficient approach,
which in certain situations and limits can be solved exactly at  arbitrary atom
densities.

In obtaining these equations, we have made the simplification that the atomic
motional state is unchanged by the scattering of light, and have ignored the corresponding
inelastic processes.
In the following sections we describe how stochastic techniques may be used to find
the polarization $\av{\Pvhat^+(\rv,t)}$, from which the electric field amplitude
$\av{\Evhat^+(\rv,t)}$ can be calculated using Eq.~\eqref{eq:MonoD}.

Equations~(\ref{eq:fullresponseequations}) are general equations including the
effects of saturation and multiple level structure.   For two-level atoms they
simplify considerably due to the fixed orientation $\pol$ of the atomic dipoles of
two-level atoms.   As a consequence, the polarization operator of the atoms is
proportional to the scalar operator $\hat{P}^+ \equiv \pol^\ast \cdot \Pvhat^+$,
and components of $\Pvhat^+$ orthogonal to $\pol$ are zero.  By making the
substitution $\Pvhat^+ \rightarrow \pol \hat{P}^+$ in
Eqs.~\eqref{eq:fullresponseequations} we obtain the two-level atom expressions
\begin{align}
{d\over dt}\,\hat{P}^+ &=
(i\bar{\Delta}-\gamma)\hat{P}^+ \nonumber \\
&+ i\xi \left(\hat\psi^\dagger_{g}\hat\psi_{g}
-\hat\psi^\dagger_{e}\hat\psi_{e}\right)
\pol^*\cdot\Dvhat^+_F \nonumber\\
&+ i\xi
\int d^3r'\,
\pol^*\cdot{\sf G}'({\bf r}-{\bf r'})\pol\,\hat\psi^\dagger_{g}\hat{P}^+(\rv')
\hat\psi_{g}\nonumber\\
& -i\xi
\int d^3r'\,
\pol^*\cdot{\sf G}'({\bf r}-{\bf r'})\pol\,\hat\psi^\dagger_{e}\hat{P}^+(\rv')
\hat\psi_{e}\label{eq:dPdt_2levelhigh}
\,, \\
{d\over dt}\,\cpsi_{g}\dpsi_{g} &= -{d\over dt}\,\cpsi_{e}\dpsi_{e}\,,
\\
&=2\gamma\cpsi_{e}\dpsi_{e} \nonumber
\\
& -i\frac{\xi}{\Dc^2}\hat{P}^-\pol^*\cdot
\Dvhat^+_F
+i\frac{\xi}{\Dc^2}\Dvhat^-_F\cdot\pol\hat{P}^+ \nonumber \\
&-i\frac{\xi}{\Dc}\int d^3r'\,\pol^*\cdot{\sf G}'({\bf r}-{\bf r'})
\pol\,\cpsi_{e}\hat{P}^+(\rv')\dpsi_{g} \nonumber\\
&+i\frac{\xi}{\Dc}\int d^3r'\,\pol\cdot{\sf G}'^*({\bf r}-{\bf r'})
\pol\,\cpsi_{g}\hat{P}^-(\rv')\dpsi_{e}\,.\label{eq:drhogdt_2levelhigh}
\end{align}

In the limit of low light intensity, the optical response given by
Eqs.~(\ref{eq:fullresponseequations}) can be significantly further simplified,
particularly so for atoms of isotropic polarizability with a single electronic ground state or for two-level atoms.  We
discuss these simplifications below, before considering the stochastic solution of
the resultant equations in Sec.~\ref{sec:lowintensity}. We subsequently extend the
stochastic treatment to the full equations in Sec.~\ref{sec:highintensity}.

\subsection{The limit of low light intensity}
\label{sec:lowlightintensity}

In the limit of low light intensity, the atoms occupy the electronic ground states,
and the change in the excited state field amplitudes is linearly proportional to the incident light
field amplitude, according to Eq.~\eqref{eq:EXF}. In order to take the weak excitation limit
in the first order of the light field amplitude $\Dvhat_F^\pm$
we therefore include in the dynamics the leading order contributions given by
terms that include at most one of the operators $\Dvhat_F^\pm$,
$\dpsi_{e\eta}$, or $\cpsi_{e\eta}$. In the first order of light intensity, the excited state
population accordingly vanishes.
Neglecting higher order terms, we obtain the equation of
motion for the polarization operator
component $\Pvhat^+_{\nu\eta}(\rv)$ from Eq.~\eqref{Pfullnew},
\begin{align}
  {d\over dt}\,\Pvhat^+_{\nu\eta} &=
  (i\bar{\Delta}_{g\nu e\eta}-\gamma)\Pvhat^+_{\nu\eta}
  + i\xi \hat\psi^\dagger_{g\nu}\hat\psi_{g\tau} {\sf P}^{\nu\eta}_{\eta\tau}
   \Dvhat^+_F \nonumber\\
  & + i\xi
  \int d^3r'\,{\sf P}^{\nu\eta}_{\eta\tau}
  {\sf G}'({\bf r}-{\bf r'})\,\hat\psi^\dagger_{g\nu}\Pvhat^+(\rv')
  \hat\psi_{g\tau}\,.
  \label{eq:CR1}
\end{align}
On the other hand, any change in the electronic ground-state coherence operators
$\cpsi_{g\nu}\dpsi_{g\eta}$ exhibits merely a phase rotation due to Zeeman
splitting. This is because any change in the ground-state amplitude
$\dpsi_{g\nu}$ beyond the phase rotation is second order in the incident light
field amplitude (via $\Ev$ and $\dpsi_{e\eta}$) according to Eq.~\eqref{eq:GRF}.

\subsubsection{Isotropic and two-level atoms}
\label{sec:formalism_lowlight_singlegroundstate}

The optical response simplifies further if we consider two special cases of atoms
with a single electronic ground level: (1) a gas of atoms whose polarizability is
isotropic with a single electronic ground state and (2) a gas of two-level atoms. An isotropic polarizability is
obtained when the atoms comprising the gas have a ground state with total
electronic and atomic angular momenta $J=0$, such as alkaline-earth-metal or
rare-earth metal atoms with zero nuclear spin. By selection rules, the excited
state must have angular momentum $J'=1$. In the absence of Zeeman splitting
between the three different excited sublevels, the atoms exhibit an isotropic
polarizability.  Two-level systems can be obtained from such atoms by choosing a
transition corresponding to an electric dipole along a desired direction.
Alternatively, a two-level system can be realized using cycling transitions in
alkali-metal atoms. In either scenario, other unwanted transitions could be
detuned out of resonance using, for example, static fields to lift the Zeeman
degeneracy.

Since only one ground state $\ket{g,\nu=0}$ is present for a gas of atoms with
isotropic polarizability, the total polarization of the system is $\Pvhat^+ =
\sum_{\eta=-1}^1 \Pvhat_{0\eta}^+$, and the relevant Clebsch-Gordan coefficients
are $\mathcal{C}_{0,\eta}^{(\sigma)} = \delta_{\eta\sigma}$.  Consistent with
Eq.~\eqref{eq:CR1}, we find the polarization evolves as~\cite{Ruostekoski1997a}
\begin{align}
  {d\over dt}\, \Pvhat^+ &=
  (i\bar{\Delta}-\gamma)\Pvhat^+
  + i\xi \hat\psi^\dagger_{g}\hat\psi_{g}
  \Dvhat^+_F \nonumber\\
  & + i\xi
  \int d^3r'\,\radKernel'({\rv}-\rv')\,\hat\psi^\dagger_{g}\Pvhat^+(\rv')
  \hat\psi_{g}\,,
  \label{eq:P_eqm_iso_body}
\end{align}
where $\bar{\Delta} = \Delta_e-\Delta_g$.

In contrast to atoms with isotropic polarizability, the fixed orientation of the
atomic dipoles of two-level atoms means that, in the low light intensity limit, we
only need to consider the equation of motion for the scalar polarisation operator
$\hat{P}^+$, which reduces in this limit to [from Eq.~\eqref{eq:dPdt_2levelhigh}]
\begin{align}
  {d\over dt}\, \hat{P}^+ &=
  (i\bar{\Delta}-\gamma)\hat{P}^+
  + i\xi \hat\psi^\dagger_{g}\hat\psi_{g}
   \pol^\ast \cdot\Dvhat^+_F \nonumber\\
  & + i\xi
  \int d^3r'\, \pol^\ast
  \cdot\radKernel'({\rv}-\rv')\pol\,\hat\psi^\dagger_{g} \hat{P}^+(\rv')
  \hat\psi_{g}\,.
  \label{eq:P_eqm_two_level}
\end{align}

In the low light intensity limit, it remains to solve
Eqs.~\eqref{eq:P_eqm_two_level},~\eqref{eq:P_eqm_iso_body}, or~\eqref{eq:CR1},
depending on the complexity of the system to be modeled.  We show in the following
section the degree to which these equations may be treated by stochastic
techniques, and that they may be solved exactly in the case of atoms with a single
electronic ground state. Later, in Sec.~\ref{sec:highintensity}, we discuss the
extension of these techniques to higher intensities, including the effects of
saturation.

\section{Stochastic classical electrodynamics simulations in the low light
intensity limit}
\label{sec:lowintensity}

\subsection{Exact solutions for atoms with a single ground level}
\label{sec:lowintensitytwolevel}

In the previous sections we introduced a quantum field-theoretical formalism for
the coupled system of atoms and light.  We will next use simple principles to
derive equations of motion for stochastic classical electrodynamics simulations of
the optical response.  For atoms with a single electronic ground state, and in the
limit of low light intensity (see
Sec.~\ref{sec:formalism_lowlight_singlegroundstate}),  these stochastic
simulations can provide exact solutions for the optical response of an ensemble of
stationary atoms that in the full  quantum field-theoretical analysis is
represented by the hierarchy of equations of motion for correlation functions
(App.~\ref{sec:monte-carlo-eval}). In contrast, as we show in the following
section, when generalizing this technique to treat atoms with multiple ground
levels we must introduce a classical approximation for virtual fluctuations
between different ground levels.

In the low light intensity limit we shall show that the atoms can be treated as
a collection of coupled linear oscillators whose positions are stochastically
sampled from the appropriate spatial distribution. For each stochastic realization
of a fixed set of atomic positions we solve the coupled electrodynamics for the
atoms and light. The optical response of the ensemble is then given by averaging
any derived quantities of interest, such as polarization, over many such
realizations~\cite{Javanainen1999a,Jenkins2012a,Javanainen2014a}.  It was first
shown in the case of 1D scalar electrodynamics in Ref.~\cite{Javanainen1999a}
that the stochastic classical electrodynamics simulations reproduce the full
quantum field-theoretical  representation for the hierarchy of equations of
motion for light-induced correlation functions [Eq.~\eqref{eq:corrFuncEqsOfM}].
Analogously, the equivalence between the classical result and the collective emission of a single-photon excitation from a cloud of $N$ two-level atoms was demonstrated in Ref.~\cite{SVI10}.
We extend the simplified model of Ref.~\cite{Javanainen1999a} to the 3D case of
the $J=0\rightarrow J'=1$ transition in App.~\ref{sec:monte-carlo-eval},
including the vector properties of the electromagnetic fields. Such coupled
dipole model simulations treating atoms as classical, linear oscillators have
been used to calculate the responses of two-level
atoms~\cite{Javanainen1999a,AkkermansEtAl2008,Jenkins2012a}, emitters with
isotropic
susceptibility~\cite{dalibardexp,Pierrat2010,Castin13,Javanainen2014a}, or
emitters with some spatial averaging over dipole orientations~\cite{pinheiro04}.

The procedure for calculating the response amounts to a Monte-Carlo integration
of Eqs.~\eqref{eq:P_eqm_iso_body} (for isotropic atoms) or
\eqref{eq:P_eqm_two_level} (for two-level atoms) in which a realization of
discrete atomic positions $\{\stochxv_1,\stochxv_2,\ldots,\stochxv_N\}$ is a set
of random variables sampled from the joint probability distribution $\vert\Psi(\rv_1,\rv_2,\ldots,\rv_N)\vert^2$, determined by the
ground-state many-body wave function $\Psi(\rv_1,\rv_2,\ldots,\rv_N)$ that
describes the center-of-mass state of the atoms.
If the atoms can be considered to be initially uncorrelated before the light enters
the medium (for instance, in a classical vapor or in an ideal Bose-Einstein condensate),
the sampling reduces to the simple procedure of
independently sampling the position of atom $i$ from the atomic density
distribution of the ensemble $\rho_1(\br_i)$ [as discussed further in
App.~\ref{sec:monte-carlo-eval}, see Eqs.~\eqref{eq:rholwave}
and~\eqref{eq:rho1factorised}]. An explicit example of the position sampling is
given later, in Sec.~\ref{sec:mott-insulator-state}, for a simple special case in
a strongly correlated quantum system.  Other examples include Fermi-Dirac
statistics, which can be modeled using a Metropolis
algorithm~\cite{Javanainen1999a}.  Quantum mechanical expectation values for the
quantities of interest are then generated from an ensemble average over many
stochastic realizations.

We will next present a simple procedure for deriving the classical electrodynamics equations
for individual realizations of atomic positions $\{\stochxv_1,\ldots,\stochxv_N\}$.
In order to obtain the equations governing the
optical response of any particular stochastic realization, we take the expectation values
of both sides of Eq.~\eqref{eq:P_eqm_iso_body}, conditioned on the $N$ atoms
being at the positions $\stochxv_1,\ldots,\stochxv_N$.
Since up until now we have been dealing with quantum field operators for the atoms,
the procedure is tantamount to
a hypothetical situation in which the atoms are physically pinned at fixed
positions $\{\stochxv_1,\ldots,\stochxv_N\}$.   The atomic density then
becomes
\begin{equation}
  \label{eq:dens_2_level}
  \quantmean{\hat{\psi}_g^\dag\hat{\psi}_g}_{\{\stochxv_1,\ldots,\stochxv_N\}} =
  \sum_j \delta(\rv-\stochxv_j)\, ,
\end{equation}
where the subscript $\{\stochxv_1,\ldots,\stochxv_N\}$ indicates the conditioning
of the expectation value.
Similarly, the polarization density operator is
\begin{equation}
  \label{eq:pol_first_qant_2_level}
  \quantmean{\Pvhat^+(\rv)}_{\{\stochxv_1,\ldots,\stochxv_N\}} = \Dc
  \sum_j \spvec{\Pc}^{(j)} \delta(\rv - \stochxv_j) \, ,
\end{equation}
where, for isotropic atoms, we define the normalized dipole amplitude
$\spvec{\Pc}^{(j)} = \sum_{\sigma}\pol_\sigma \Pc_{0\sigma}^{(j)}$ of atom $j$,
conditioned on atoms being located at positions $\stochxv_1,\ldots,\stochxv_N$.  Similarly,
for two-level atoms, we define the conditional normalized dipole amplitude
$\spvec{\Pc}^{(j)} = \pol \Pc^{(j)}$.

In the following, we will proceed explicitly with the derivation of the stochastic
equations of motion for the isotropic case $J=0\rightarrow J'=1$. The expectation value for the density-polarization
operator appearing
under Eq.~\eqref{eq:P_eqm_iso_body} is
\begin{eqnarray}
  \lefteqn{\quantmean{\hat\psi^\dagger_{g}(\rv)\Pvhat^+(\rv')\hat\psi_{g}(\rv)}_{\{\stochxv_1,\ldots,\stochxv_N\}}
  } \nonumber\\
  &=&
  \Dc\sum_j\sum_{l\neq j} \spvec{\Pc}^{(l)}
  \delta(\rv-\stochxv_j)\delta(\rv'-\stochxv_l) \, .
  \label{eq:two_body_corr}
\end{eqnarray}
Here we have explicitly avoided double counting of any atom
by excluding the atom $j$ in the summation over $l$. The benefit of
arranging all the atomic operators into normal order now becomes
obvious.
The normal ordering corresponds to a simple procedure of classical
counting of atoms, in an analogy of photon counting theory in photon
detection~\cite{Glauber1963a,Glauber1963b,Kelley1964a}.

Substituting Eqs.~\eqref{eq:dens_2_level},
\eqref{eq:pol_first_qant_2_level} and \eqref{eq:two_body_corr} into
Eq.~\eqref{eq:P_eqm_iso_body}, we find that for isotropic atoms
\begin{widetext}
  \begin{align}
    {d\over dt} & \sum_{j} \spvec{\Pc}^{(j)} \, \delta(\rv-\stochxv_j) =
    (i\bar{\Delta}-\gamma)\sum_{j} \spvec{\Pc}^{(j)}  \, \delta(\rv-\stochxv_{j})
    + i{\xi\over \Dc}\,\sum_{j}  {\Dv}^+_F \, \delta(\rv-\stochxv_{j})  \nonumber\\
    & +i\xi \sum_{j} \sum_{l\neq j} \int d^3r'\, {\sf G}'({\bf
      r}-{\bf r'})  \spvec{\Pc}^{(l)} \delta(\rv-\stochxv_j)
    \delta(\rv'-\stochxv_l)\,, \label{eq:Pol_single_atom_isotropic_intermediate}
  \end{align}
\end{widetext}
where we have taken the driving field ${\Dvhat}^+_F$ to be in a coherent state, such
that under expectation values it provides a multiplicative classical coherent
field ${\Dv}^+_F$.
Multiplying both sides of
Eq.~\eqref{eq:Pol_single_atom_isotropic_intermediate} by
$\delta_{j,j'}$, interchanging indices $j$ and $j'$, and integrating over $\rv$, we find the dynamics of
the individual atomic dipole moments satisfy the coupled equations
\begin{align}
  {d\over dt} & \spvec{\Pc}^{(j)} = (i\bar{\Delta}-\gamma) \spvec{\Pc}^{(j)}
  + i{\xi\over \Dc}\,  {\bf D}^+_F(\stochxv_j) \nonumber\\
  & +i\xi \sum_{l\neq j}
  {\sf G}'(\stochxv_j-\stochxv_l) \spvec{\Pc}^{(l)} \, ,
  \label{eq:Pol_single_atom_J0_J1}
\end{align}
which may be used to study the time-dynamics of the polarization, e.g., by short resonant pulses.
We may also obtain the steady-state solution to Eq.~\eqref{eq:Pol_single_atom_J0_J1}
\begin{equation}
  \Pcv^{(j)}= \frac{\alpha}{\Dc}\,
  \Dv^+_F(\stochxv_j) + \alpha\sum_{l\neq j}
  {\sf G}'(\stochxv_j-\stochxv_l)\Pcv^{(l)}\,,
\label{eq:stestapol}
\end{equation}
where the polarizability of the atom $\alpha$ is in this case isotropic, and defined by
\begin{equation}
  \alpha=-{\Dc^2\over \hbar\epsilon_0(\bar{\Delta}+i\gamma)}\,.
  \label{eq:polar}
\end{equation}
Both Eqs.~\eqref{eq:Pol_single_atom_J0_J1} and~\eqref{eq:stestapol} represent the coupled equations
for atomic excitations in the presence of a driving incident light. The atoms
are located at discrete
positions $\{\stochxv_1,\stochxv_2,\ldots,\stochxv_N\}$ and the interactions between the atoms are mediated by the scattered light field.  As a
consequence $\spvec{\Pc}^{(j)}$ depends not only on $\stochxv_j$, but also on the
position of all other atoms.

Each atom acts as a source of scattered dipole radiation providing a scattered
field for (the positive frequency component) of the electric field $\eo {\bf E}_S^{(l)}(\rv)  = {\sf
G}'(\rv-\stochxv_l)\Dc \spvec{\Pc}^{(l)}$, where the propagator is defined in
Eq.~\eqref{eq:Gprime} and gives the familiar expression for the electric field
radiated by a dipole $\Dc\spvec{\Pc}^{(l)}$ situated at point $\stochxv_l$. Each atom $j$ at position $\stochxv_j$ in Eqs.~\eqref{eq:Pol_single_atom_J0_J1} and~\eqref{eq:stestapol} is therefore driven by the field
$\bE_{\rm ext}^+(\stochxv_j)$ that is the sum of the incident field
$\Dv^+_F(\stochxv_j)$ and the fields scattered from all $N-1$ other atoms in the system
\begin{equation}
  \epsilon_0 \bE_{\rm ext}^+(\stochxv_j) = \Dv^+_F(\stochxv_j)+\sum_{l\neq j}
  \epsilon_0\bE^{(l)}_S(\stochxv_j)\,.
\end{equation}
Focusing on the steady-state response~\eqref{eq:stestapol}, the scattered field then induces the polarization at the atom $j$ that is
proportional to the polarizability $\alpha$ [Eq.~\eqref{eq:polar}]
$\Dc \spvec{\Pc}^{(j)} = \alpha\epsilon_0   {\bf E}_{\rm
  ext}^+(\stochxv_j)$.
Owing to the isotropic response of the atoms in case of the $J=0\rightarrow J'=1$ transition,
the total electric field in the steady-state case~\eqref{eq:stestapol} may be solved from
\begin{equation}
  \epsilon_0 \bE_{\rm ext}^+(\stochxv_j) = \Dv^+_F(\stochxv_j) + \epsilon_0   \alpha \sum_{l\neq j}
  {\sf G}'(\rv-\stochxv_l)  {\bf E}_{\rm ext}^+(\stochxv_l) \,.
\label{eq:linearisoexc}
\end{equation}
This linear set of equations can be solved to give the total electric field amplitude everywhere in space
\beq
\epsilon_0\bE^+(\rv) = \Dv^+_F(\rv) + \epsilon_0 \alpha \sum_{j}  {\sf
G}'(\rv-\stochxv_j){\bf E}_{\rm ext}^+(\stochxv_j)\,.
\label{eq:E_tot_singlelevel}
\eeq
A convenient feature of the isotropic $J=0\rightarrow J'=1$ transition
is that all the internal properties of the atoms in
Eq.~\eqref{eq:linearisoexc} are encapsulated in a single scalar
quantity $\alpha$.

The response of the system at low light intensities for a
single realization of atomic positions is equivalent to the response of
a model of coupled harmonic oscillators at positions $\stochxv_1,\ldots,\stochxv_N$.  The
oscillators have dipole matrix elements $\Dc$, detuning $\bar{\Delta}$, and spontaneous
linewidth $\gamma$ defined in Eq.~\eqref{eq:spontlinewidth}.

In order to obtain the steady-state solution for the optical response of the
atomic ensemble where the positions of the atoms are distributed according to some
joint probability distribution $P(\rv_1, \rv_2,\ldots,\rv_N)$, one then
stochastically samples the atomic positions. For each stochastic realization of a
set of discrete atomic positions $\{\stochxv_1,\stochxv_2,\ldots,\stochxv_N\}$,
one solves Eq.~\eqref{eq:linearisoexc} for the electric field, or equivalently
Eq.~\eqref{eq:stestapol}. With that solution, one calculates the observables of
interest for each realization. The expectation value of those observables
corresponds to the average over sufficiently many stochastic realizations.
Dynamical calculations proceed analogously, calculating dynamical trajectories
from Eq.~\eqref{eq:Pol_single_atom_J0_J1} for individual realizations of atomic
positions, and averaging over many trajectories.

Following the arguments of Ref.~\cite{Javanainen1999a}, we show in App.~\ref{sec:monte-carlo-eval} that this sampling
procedure generates the correct dynamics for many-body
polarization-density correlation functions in the low light intensity limit.  We note
that, while we appear to neglect correlations between atoms in a single realization
by using Eq.~\eqref{eq:two_body_corr}, App.~\ref{sec:monte-carlo-eval}
demonstrates the subsequent averaging over many
stochastic realizations includes all relevant correlations for low light intensity
coherent scattering.

While we focus on cold, homogeneously broadened ensembles of atoms, we note that
thermal effects can also be approximated in this treatment. Thermal motion of
the atoms introduces Doppler broadening of the resonance frequencies. This
inhomogeneous broadening mechanism is accounted for by additionally
stochastically sampling the resonance frequency detunings of each atom
$\bar{\Delta}^{(j)}$ to reflect the distribution of thermal atomic
velocities~\cite{Javanainen2014a}.

The two-level case can be treated analogously to the isotropic case.
In contrast to atoms with isotropic polarizability, the atomic dipoles
of two-level atoms have a fixed orientation $\pol$.  As a consequence, the dipole moment of
the atoms is proportional to the scalar $\Pc^{(j)} \equiv \pol^\ast \cdot
\spvec{\Pc}^{(j)}$, and components of $\spvec{\Pc}^{(j)}$ orthogonal
to $\pol$ are zero.
By making the substitution $\spvec{\Pc}^{(j)} \rightarrow \pol
\Pc^{(j)}$ into Eq.~\eqref{eq:Pol_single_atom_J0_J1}, we find that for a
stochastic realization of two-level atomic positions, the dipole
amplitudes evolve as
\begin{align}
  {d\over dt} & \Pc^{(j)} = (i\bar{\Delta}-\gamma) \Pc^{(j)}
  + i{\xi\over \Dc}\, \pol^\ast \cdot  {\bf D}^+_F(\stochxv_j) \nonumber\\
  & +i\xi \sum_{l\neq j}
  \pol^{\ast}\cdot{\sf G}'(\stochxv_j-\stochxv_l)  \pol \,\Pc^{(l)} \, .
  \label{eq:Pol_single_atom_two_level}
\end{align}
Averaging over many such realizations generates the dynamics of the expectation
value of the polarization, from which quantities of interest may be calculated.
Alternatively, analogous arguments to those above can be used to directly solve
for the steady state electric field.

\subsection{Approximate solutions for multilevel atoms}
\label{sec:lowintensitymultilevel}

In common experimental situations~\cite{Pellegrino2014a},  the electronic ground state of the
atoms may have nonzero angular momentum and multiple internal transitions
participate in the scattering processes.  In this section we consider the
generalization of the stochastic classical electrodynamics simulations of
Sec.~\ref{sec:lowintensitytwolevel} to such cases.
In contrast to single ground level atoms, the systems consisting of atoms with multiple ground levels
support virtual fluctuations between Zeeman ground
levels that can become correlated with the atomic polarization, even in the low
light intensity limit. This added complexity manifests itself in the quantum
field-theoretical representation for the coupled system of atoms and light that
leads to the hierarchy of equations of motion for $\ell^{\mathrm{th}}$ order
correlation functions ($\ell = 1,\ldots,N$) that fully describe the ensemble's
optical response [see
Eqs.~\eqref{eq:DensityCorrFuncDef_general}-\eqref{eq:corrFuncEqsOfM_general}].  As
shown in App.~\ref{sec:monte-carlo-eval}, rather than considering one
vector-valued correlation function for each level of the dynamic hierarchy for
isotropic atoms, the number of correlation functions involved in the optical
response of multilevel atoms scales exponentially with the number of atoms in the
system.

Physically, for single ground level atoms at the limit of low light intensity,
the excitation of each atom can be described as a superposition of linear
polarization amplitudes. There is therefore no additional internal level
dynamics beyond the spatial positions of the atoms that could become correlated
by the scattered light. A numerical treatment of the light-induced position
correlations between the atoms is consequently sufficient to obtain the complete
solution to the problem. However, for atoms possessing multiple ground levels
the light may also generate nonclassical internal level correlations between
different atoms. As a consequence, the classical electrodynamics simulations of
charged harmonic oscillators at stochastically sampled positions that exactly
reproduce the hierarchy for isotropic and two-level atoms (see
Sec.~\ref{sec:lowintensitytwolevel}) are not able to capture quantum fluctuations
that may potentially arise in the case of multiple ground levels.

In this section, we show how to generalize the stochastic classical electrodynamics
simulations of Sec.~\ref{sec:lowintensitytwolevel} to approximate the cooperative
response of an ensemble of atoms with multiple electronic ground states. We
introduce a classical approximation where we include the recurrent scattering
processes between the classical dipoles corresponding to the different atomic
transitions, but nonclassical higher order correlations between internal levels that involve
fluctuations between ground-state coherences of different levels and the
polarization are factorized.
We emphasize, however, that the \emph{single}-particle expectation values of the  ground state populations and
coherences, themselves, are unaffected
by these nonclassical correlation effects to first order in the electric field amplitude.
An analogous  factorization approximation of higher
order internal atomic level correlations will also be employed in
Sec.~\ref{sec:highintensity} when we  show how to simulate an
ensemble's cooperative response outside the low light intensity limit.
The classical electrodynamics
approximation for atoms with multiple ground levels   leads to a treatment where
the  atomic Zeeman levels, as well as the atomic positions, are stochastically
sampled. This was first outlined and employed in Ref.~\cite{Pellegrino2014a} to
model the experiments presented in that reference.  We later show how to improve the
approximation to account for pairwise correlations between virtual fluctuations of
Zeeman states and atomic dipoles in Sec.~\ref{sec:incorp-two-body}.

As with isotropic and two-level atoms, we determine the optical response of the gas
by stochastically sampling the positions of the atoms
$\{\stochxv_1,\ldots,\stochxv_N\}$, and then determine the optical response for each
such realization. Following the same general procedure established in
Sec.~\ref{sec:lowintensitytwolevel}, for each realization of atomic positions, we
consider the hypothetical situation where atoms are pinned to the points
$\{\stochxv_1,\ldots,\stochxv_N\}$. The dynamics of the atomic dipoles for each
realization are then found by taking the expectation value of Eq.~\eqref{eq:CR1}
conditioned on atoms being located at those points.
For notational simplicity, we focus here on the common experimental situation where the atoms are
initially (before the incident light enters the atomic gas) assumed to be in an incoherent mixture of ground levels $\ket{g,\nu}$, such
that  $\av{\cpsi_{g\nu}\dpsi_{g\tau}}=0$ if $\nu \neq \tau$. The treatment can be generalized also
to the case where the initial Zeeman coherences between the different ground levels $\nu \neq \tau$ are nonvanishing
and this will be briefly discussed towards the end of this section.
In the limit of low light intensity, an initially incoherent mixture remains so at all times, and the
relative populations of the different levels are invariant.

The ground-level densities conditioned on a single realization of discrete atomic
positions can conveniently be represented as
\begin{equation}
  \< \hat\psi^\dagger_{g\nu}\hat\psi_{g\nu} \>_{\{\stochxv_1,\ldots,\stochxv_N\}} =
  \sum_j f_\nu^{(j)} \delta(\rv-\stochxv_j),
\label{eq:definediscretef}
\end{equation}
with relative occupations of the different ground levels $f_\nu^{(j)}$ ($0 \le
f_\nu^{(j)} \le 1$ and $\sum_\nu f_{\nu}^{(j)} = 1$), at position $\stochxv_j$.
In the limit of low light intensity these single-atom ground-state populations are constant and unaffected
by light to first order in the electric field amplitude (see Sec.~\ref{sec:lowlightintensity}).
Analogous to Eq.~\eqref{eq:pol_first_qant_2_level}, we similarly obtain for a
specific realization of the atomic positions the polarization
\begin{align}
  \av{\Pvhat_{\nu\eta}^+}_{\{\stochxv_1,\ldots,\stochxv_N\}} & =
  \Dc \sum_{\sigma}\pol_{\sigma} {\cal C}_{\nu,\eta}^{(\sigma)}
    \nonumber \\
    &  \times \sum_j f_\nu^{(j)}\Pc_{\nu\eta}^{(j)}(t)\, \delta(\rv-\stochxv_j),
\label{eq:definediscretePol}
\end{align}
where $\Pc_{\nu\eta}^{(j)}(t)$ denote the dynamic excitation amplitudes. The electric dipole
for the atomic transition $|g,\nu\rangle\rightarrow |e,\eta\rangle$ of atom $j$ is then given by
$\Dc\pol_{\eta-\nu}
  \mathcal{C}_{\nu,\eta}^{(\eta-\nu)} f_\nu^{(j)} \Pc_{\nu\eta}^{(j)}$.

The expectation value for the density-polarization operator [appearing in the last
term of Eq.~\eqref{eq:CR1}] involves the two-body correlation
functions
\begin{align}
  &\quantmean{\hat\psi^\dagger_{g\nu}(\rv) \psi^{\dagger}_{g\mu}(\rv')
    \hat{\psi}_{e\eta}(\rv')
    \hat\psi_{g\tau}(\rv)}_{\{\stochxv_1,\ldots,\stochxv_N\}}
  \nonumber \\
  &=
  \sum_{j}\sum_{l\ne j}
  \Pc_{\nu\tau;\mu\eta}^{(j;l)}
  \delta(\rv-{\stochxv}_j)\delta(\rv'-\stochxv_l) \, .
  \label{eq:multi_level_Stoch_2_body_corr_def}
\end{align}
Here $\Pc_{\nu\tau;\mu\eta}^{(j;l)}$ denote two-atom internal level correlation
functions. They characterize correlations involving a Zeeman coherence between the
ground levels $\nu$ and $\tau$ of atom $j$ and the polarization amplitude for the
transition $|g,\mu\rangle\rightarrow |e,\eta\rangle$ of another atom $l\ne j$. It
might, at first, appear unnecessary to include such pair correlations that consists of
a Zeeman coherence between \emph{different} ground levels $\nu$ and $\tau$. We assumed
an incoherent mixture for which $\< \hat\psi^\dagger_{g\nu}\hat\psi_{g\tau} \>=0$
whenever $\nu\neq\tau$ -- which then in the limit of low light intensity remains valid at all
times.  However, while no single particle Zeeman coherences can be generated by
optical scattering in an incoherent mixture in the low light intensity limit, we will
show that light scattering \emph{can} generate nonzero two-atom correlations
$\Pc_{\nu\tau;\mu\eta}^{(j;l)}$, and we discuss the implications of this below.

Using these definitions by substituting Eqs.~\eqref{eq:definediscretef},
\eqref{eq:definediscretePol}, and \eqref{eq:multi_level_Stoch_2_body_corr_def}
into Eq.~\eqref{eq:CR1}, we obtain (for
$f_{\nu}^{(j)} \ne 0$)
\begin{widetext}
\begin{align}
  {d\over dt} & \sum_j \Pc_{\nu\eta}^{(j)} \, \delta(\rv-\stochxv_j) =
  [i\bar{\Delta}_{g\nu e\eta}-\gamma]\sum_j
  \Pc_{\nu\eta}^{(j)}  \, \delta(\rv-\stochxv_j)
  + i\frac{\xi}{\Dc}\,\sum_j  {\cal C}_{\nu,\eta}^{(\sigma)} \pol_{\sigma}^\ast \cdot
  {\bf D}^+_F \, \delta(\rv-\stochxv_j)  \nonumber\\
  & +i\xi \sum_j \sum_{l\neq j} \int d^3r'\, {\cal
    C}_{\tau,\eta}^{(\sigma)} \pol_{\sigma}^\ast  \cdot
  {\sf G}'(\rv-\rv')\pol_{\varsigma}\mathcal{C}_{\mu,\zeta}^{(\varsigma)}
  \frac{\Pc_{\nu\tau;\mu\zeta}^{(j;l)}}{f_\nu^{(j)}}\delta(\rv-\stochxv_j)
  \delta(\rv'-\stochxv_l)\, \textrm{,}
  \label{eq:pol_corr_eqm_multi-level}
\end{align}
\end{widetext}
  where $\bar{\Delta}_{g\nu e\eta} = \Delta_{e\eta}-\Delta_{g\nu}$, and
  repeated indices $\tau$, $\mu$, $\zeta$, $\sigma$ and $\varsigma$ are
  summed over.

As in Sec.~\ref{sec:lowintensitytwolevel}, we can simplify
Eq.~\eqref{eq:pol_corr_eqm_multi-level} by multiplying both sides by
$\delta_{j,j'}$, interchanging the indices $j$ and $j'$, and
integrating over $\rv$, thus obtaining
\begin{align}
  {d\over dt} &\Pc_{\nu\eta}^{(j)} = (i\bar{\Delta}_{g\nu e\eta}-\gamma)\Pc_{\nu\eta}^{(j)}
  + i\frac{\xi}{\Dc}\,  \mathcal{C}_{\nu,\eta}^{(\sigma)} \pol_{\sigma}^\ast\cdot
  \Dv^+_F(\stochxv_j) \nonumber\\
  & +i\xi \sum_{l\neq j} {\cal C}_{\tau,\eta}^{(\sigma)}\pol_{\sigma}^\ast\cdot
  \radKernel'(\stochxv_j-\stochxv_l)  \pol_{\varsigma}\mathcal{C}_{\mu,\zeta}^{(\varsigma)}
  \frac{\Pc_{\nu\tau;\mu\zeta}^{(j;l)}}{f_\nu^{(j)}}\, \textrm{,}
  \label{eq:Pcv_multi_level_dynamics}
\end{align}
In the absence of magnetic Zeeman splitting ($B=0$),  the
steady state polarization amplitudes for a specific atom
$j$ become
\begin{align}
  \Pc_{\nu\eta}^{(j)} &=
  \frac{\alpha_{\nu\eta}}{\Dc}\, \mathcal{C}_{\nu,\eta}^{(\sigma)} \pol_{\sigma}^\ast\cdot
  \Dv^+_F(\stochxv_j) \nonumber\\
  & + \alpha_{\nu\eta} \sum_{l\neq j}  {\cal C}_{\tau,\eta}^{(\sigma)} \pol_{\sigma}^\ast\cdot
  {\sf G}'(\stochxv_j-\stochxv_l) \pol_{\varsigma} \mathcal{C}_{\mu,\zeta}^{(\varsigma)}
  \frac{\Pc_{\nu\tau;\mu\zeta}^{(j;l)}}{f_\nu^{(j)}}\,.
  \label{eq:Pvc_multi_level_steady_state}
\end{align}
In Eqs.~\eqref{eq:Pcv_multi_level_dynamics} and
\eqref{eq:Pvc_multi_level_steady_state}, repeated indices $\tau$, $\mu$,
$\zeta$, $\sigma$, and $\varsigma$ are summed over.
The atomic polarizability is now anisotropic
\beq
\alpha_{\nu\eta}=-{\Dc^2\over \hbar \eo (\bar{\Delta}_{g\nu
  e\eta}+i\gamma)}\,.
  \eeq

We have now removed the spatial correlations of the atoms by considering the atoms at fixed positions.
In the simulations this represents an individual stochastic realization when the atomic configuration has been sampled from the joint
probability distribution of the atomic positions. However, due to the multiple internal ground levels, correlations between the levels may still exist.
Generally, to calculate the optical response, one must therefore determine the dynamics of the
two-atom internal level correlations $\Pc_{\nu\tau;\mu\zeta}^{(j;l)}$. We will see
later in Sec.~\ref{sec:incorp-two-body} that the internal level pair correlations
depend on three-atom internal level correlations, and so on. The equations of motion
for these internal level correlation functions may be derived from the quantum
field-theoretical hierarchy of equations of motion for multilevel density-polarization
correlation functions presented in App.~\ref{sec:monte-carlo-eval} [see
Eqs.~\eqref{eq:DensityCorrFuncDef_general}-\eqref{eq:corrFuncEqsOfM_general}] when the
atoms are pinned at the stochastically sampled points
$\{\stochxv_1,\ldots,\stochxv_N\}$. In contrast to the single ground level case (as
discussed in Sec.~\ref{sec:lowintensitytwolevel}), for multiple ground levels the full
field-theoretical description [Eq.~\eqref{eq:corrFuncEqsOfM_general}] cannot be solved
exactly by tracking only single-atom internal level correlations for each realization
of atomic positions. Rather, $\Pc_{\nu\tau;\mu\zeta}^{(j;l)}$ does not factorize into
single-atom internal level correlations and describes pair correlations between
internal levels, involving virtual fluctuations between ground-state coherences of
different levels and the polarization. We will briefly discuss the dynamics of
$\Pc_{\nu\tau;\mu\zeta}^{(j;l)}$ in Sec.~\ref{sec:incorp-two-body}

Here we introduce a computationally efficient classical approximation where the virtual fluctuations between ground-state coherences of
different levels and the polarization are neglected. We assume that
the correlations
$\Pc_{\nu\tau;\mu\zeta}^{(j;l)}$ for
the atoms $j$ and $l$ factorize,
\begin{equation}
  \Pc_{\nu\tau;\mu\eta}^{(j;l)} \simeq
  \delta_{\nu\tau}f_\nu^{(j)}f_\mu^{(l)}\Pc_{\mu\eta}^{(l)} \, {,}
  \label{eq:multi_level_P2_simple}
\end{equation}
where the Kronecker delta $\delta_{\nu\tau}$ appears due to the assumption of an incoherent mixture
of ground levels.
With this simplification, Eqs.~\eqref{eq:Pcv_multi_level_dynamics} and
\eqref{eq:Pvc_multi_level_steady_state} each become a closed set of
linear equations resembling the electrodynamic response of a set of
classical linear oscillators. These can be solved for each set of fixed
atomic positions. Ensemble-averaging over many stochastic realizations then generates
light-induced correlations between the atoms.

For a single realization, the temporal evolution and the steady state of the atomic polarization
can be inferred from Eqs.~\eqref{eq:Pcv_multi_level_dynamics} and
\eqref{eq:Pvc_multi_level_steady_state} [with
the approximation of Eq.~\eqref{eq:multi_level_P2_simple}], respectively.
For each occupied Zeeman ground state (i.e., levels
$\ket{g,\nu}$ for which $f_\nu^{(j)} \ne 0$), the amplitudes
$\Pc_{\nu,\nu+\sigma}^{(j)}$ ($\sigma = 0,\pm 1$) potentially
contribute to the dipole moment of atom $j$.
At first glance, one might expect that for every atom, one would have
to track three amplitudes for every occupied Zeeman ground level.
A numerically efficient alternative to this approach, however, is to
stochastically sample ground level occupations for every realization of atomic
positions.
Specifically, for each realization, one randomly samples the atomic positions
$\{\stochxv_1,\ldots,\stochxv_N\}$, drawn
from the many-body joint probability distribution. Then,
for each atom $j=1,\ldots,N$, one stochastically samples the atom's
initial ground level $\ket{g,M_j}$, where the probability of choosing
the Zeeman state $\{M_1,\ldots,M_N\}$ is governed by the
populations $f_\nu^{(j)}$, which can be obtained from the initial
many-body distribution.
For each stochastic realization,
each atom is then described by only three amplitudes
$\Pc_{M_j,M_j+\sigma}^{(j)}$ ($\sigma = 0, \pm 1$), which have the
steady-state solutions (with the repeated indices $\sigma$, $\varsigma$, and $\zeta$ summed
over)
\begin{align}
  &\lefteqn{\Pc_{M_j\eta}^{(j)} =
    {\alpha_{M_j\eta}\over \Dc}\, {\cal C}_{M_j ,\eta}^{(\sigma)} \pol_{\sigma}^\ast\cdot
    {\bf D}^+_F(\stochxv_j)  }\nonumber\\
  &+ \alpha_{M_j\eta} \sum_{l\neq j}  {\cal C}_{M_j,\eta}^{(\sigma)}
  \pol_{\sigma}^\ast\cdot{\sf G}'(\stochxv_j-\stochxv_l)  \pol_{\varsigma}{\cal C}_{M_l,\zeta}^{(\varsigma)}
  \Pc_{M_l\zeta}^{(l)}\,.
  \label{eq:multi_realization_steady_state}
\end{align}

From the solution of the steady-state or time-dependent amplitudes for
each realization, one can calculate the scattered electric field and
other related observables.
We can then write the
total electric field as the sum of the incident field and the fields
scattered from each atom in the ensemble, which for a single
realization of atomic positions and Zeeman levels is given by
\begin{equation}
  \label{eq:E_tot_multi_level}
  \epsilon_0\Ev(\rv,t) = \Dv_{F}^+(\rv,t) + \Dc \sum_j \radKernel'(\rv
  - \stochxv_j)  \pol_\sigma \mathcal{C}_{M_j,\eta}^{(\sigma)}
  \Pc_{M_j\eta}^{(j)} \, \textrm{,}
\end{equation}
where repeated indices $\sigma$ and $\eta$ are summed over. [The single ground level
analogue of Eq.~\eqref{eq:E_tot_multi_level} for a $J=0\rightarrow J'=1$ transition
is given by Eq.~\eqref{eq:E_tot_singlelevel}.]
Subsequent averaging over a
large enough ensemble of realizations of positions and Zeeman
states provides the expectation values of the scattered electric field and
other observables of interest.
The multilevel classical electrodynamics simulations have recently been implemented to study the pulsed excitations of $^{87}$Rb atoms in the $F=2$ ground-state manifold
for fluorescence and coherent light transmission~\cite{Pellegrino2014a,Jennewein_trans,Jenkins_thermshift}.

Though the classical approximation of Eq.~\eqref{eq:multi_level_P2_simple}
neglects the quantum virtual fluctuations between different Zeeman ground
levels, as mentioned earlier, averaging over many stochastic realizations of
positions and Zeeman states still produces correlations between the
polarization, density, and atomic Zeeman states. In particular, as in
Sec.~\ref{sec:lowintensitytwolevel}, recurrent scattering processes are included
in this treatment.  However, in contrast to the case of a single ground level
system, due to the approximation of Eq.~\eqref{eq:multi_level_P2_simple} we no
longer exactly reproduce the dynamics of the full quantum field-theoretical
representation of the optical response that is expressed by the hierarchy of
equations for the correlation functions given in
App.~\ref{sec:monte-carlo-eval}.  A similar approximation for two-level atoms
outside the low light intensity limit will be discussed in
Sec.~\ref{sec:HighIntensityTwoLevel}.

We have focused here on describing the response of atomic ensembles that initially -- before the light enters the sample -- are prepared in
an incoherent mixture of ground levels. Ground level coherences could be
included by replacing Eq.~\eqref{eq:definediscretef} by the more general
expression
\begin{equation}
  \< \hat\psi^\dagger_{g\nu}\hat\psi_{g\tau}
  \>_{\{\stochxv_1,\ldots,\stochxv_N\}} = \sum_j
  \varrho_{\nu\tau}^{(j)}\delta(\rv-\stochxv_j).
\end{equation}
Here $\varrho_{\nu\tau}^{(j)}$ describe coherences between internal ground
levels $\nu$ and $\tau$ for atom $j$ in a single realisation of atomic
positions.  In the limit of low light intensity, such coherences are invariant
with the exception of a phase rotation resulting from any energy splitting of
the ground levels.  An analogous treatment to that presented above then leads to
generalized equations of motion for a single stochastic realization.  Later, in
Sec.~\ref{sec:HighIntensityMultiLevel}, we show explicitly how ground level
coherences may be taken into account outside of the low light intensity limit,
where one allows for saturation and dynamics between all Zeeman levels.

\subsubsection{Incorporating virtual Zeeman fluctuations}
\label{sec:incorp-two-body}

The stochastic technique described above removes the spatial correlations in
each individual stochastic realization and the electrodynamics can be solved for
a fixed set of atomic positions. Generally, in the presence of multiple internal
ground levels, there is additional internal-level dynamics beyond the spatial
positions of the atoms that can in principle lead to nonclassical correlations
between the different atoms mediated by the scattered light. The classical approximation,
we introduced in the previous section, assumes that the two-atom internal level
correlation functions $\Pc_{\nu\tau;\mu\eta}^{(j;l)}$ can be factorized
according to Eq.~\eqref{eq:multi_level_P2_simple}, leading to a coupled
multilevel dipole model. In order to incorporate quantum effects beyond the
classical model, we must track the virtual fluctuations between different ground
Zeeman states in $\Pc_{\nu\tau;\mu\eta}^{(j;l)}$ in each realization of atomic
positions. We now discuss how to track this quantity, accounting for the
possibility that correlations may form between these virtual fluctuations and
atomic dipoles.

We treat the two-atom internal level correlation functions on which the single-atom amplitudes
depend [see Eq.~\eqref{eq:Pcv_multi_level_dynamics}] in a way similar to
the method we employed to derive the
dynamics for $\Pc_{\nu\eta}^{(j)}$.
Using Eqs.~\eqref{newpsie} and
\eqref{eq:dpsigdt}, one can write the equation of
motion, to first order in the electric field, of the products of field
operators appearing in Eq.~\eqref{eq:CR1}.
For a general initial many-body atomic wave function, the
polarization can be found by solving the hierarchy of equations of motion for the
correlation functions
presented in App.~\ref{sec:monte-carlo-eval}
\cite{Ruostekoski1997a}.
For a specific realization of atomic positions, one expresses the
correlation functions subject to the atoms being at those positions
$\{\stochxv_1,\ldots,\stochxv_N\}$.
Doing so results in a  two-atom ground state correlation function
that interacts with the driving field.
We write  the ground-state pair correlation
function
\begin{widetext}
  \begin{equation}
     \quantmean{ \hat{\psi}_{g\nu}^{\dag}\left(\rv_1\right)
       \hat{\psi}_{g\mu}^{\dag}\left(\rv_2\right)
       \hat{\psi}_{g\eta}\left(\rv_2\right)
       \hat{\psi}_{g\tau}\left(\rv_1\right)
     }_{\left\{ \stochxv_{1},\ldots,\stochxv_{N}\right\}  }  =\sum_{j_{1}} \sum_{j_{2}\neq j_{1}}
     \varrho_{\nu\tau,\mu\eta}^{\left(j_{1},j_{2}\right)}
     \delta\left(\rv_1-\stochxv_{1}\right)
     \delta\left(\rv_2-\stochxv_2\right) \, \textrm{,}\label{eq:internalpair}
   \end{equation}
   where $\varrho_{\nu\tau,\mu\eta}^{\left(j_{1},j_{2}\right)}$
   is the internal ground-level pair correlation function determined by the initial many-body
   wavefunction before the incident light enters the gas.
 (To first order in the electric field amplitude,
  $\varrho_{\nu\tau,\mu\eta}^{\left(j_{1},j_{2}\right)}$
  is invariant up to a phase rotation caused by any Zeeman
  splitting of the ground level.)
  We express the three-atom correlation function resulting from light
  scattering from atoms not at $\rv_1$ or $\rv_2$ as
  \begin{align}
    & \quantmean{ \hat{\psi}_{g\nu_1}^{\dag}\left(\rv_1\right)
      \hat{\psi}_{g\nu_2}^{\dag}\left(\rv_2\right)
      \hat{\psi}_{g\nu_{3}}^{\dag}\left(\rv_3\right)
      \hat{\psi}_{e\eta_3}\left(\rv_3\right)
      \hat{\psi}_{g\tau_2}\left(\rv_2\right)
      \hat{\psi}_{g\tau_1}\left(\rv_1\right) }_{\left\{\stochxv_1
        ,\ldots,\stochxv_{N}\right\}} \nonumber\\
    & = \sum_{(j_1,j_2,j_{3})}
    \Pc_{\nu_1\tau_1,\nu_2\tau_2;\nu_{3}\eta_3}^{\left(j_{1},j_{2};j_{3}\right)}
    \delta\left(\rv_1-\stochxv_1\right)
    \delta\left(\rv_2-\stochxv_2\right)
    \delta\left(\rv_3-\stochxv_3\right) \, \textrm{.}
  \end{align}
\end{widetext}
where
$\Pc_{\nu_1\tau_1,\nu_2\tau_2;\nu_3\eta_3}^{\left(j_{1},j_{2};j_{3}\right)}$
is the three-atom internal level correlation function involving ground state Zeeman coherences of atoms
$j_1$ and $j_2$ and the dipole amplitude of atom $j_3$, at the positions $\stochxv_1$, $\stochxv_2$, and $\stochxv_3$, respectively.
The summation is over all permutations of distinct indices
$(j_1,j_2,j_3)$ in the set $\{j=1\ldots N\}$.

Here, as before, we assume the atoms are initially,  in the absence of light,  in an incoherent
mixture of Zeeman levels. Furthermore,
for simplicity, in the following we will consider an atomic system that does not exhibit correlations in the absence of the incident light,
so that we can write  in Eq.~\eqref{eq:internalpair} the correlation
function
    \begin{equation}
 \varrho_{\nu\tau,\mu\eta}^{\left(j_{1},j_{2}\right)} =    f_\nu^{(j_1)} f_\mu^{(j_2)}\delta_{\nu\tau}\delta_{\mu\eta} \, \textrm{.}
  \end{equation}

For a specific realization of atomic positions, the two-atom
amplitudes evolve according to [Eq.~\eqref{eq:corrFuncEqsOfM_general}]
\begin{align}
  & \left[ \frac{d}{dt}-i\left( \bar{\Delta}_{g\mu e\eta} +
  \bar{\Delta}_{g\nu g\tau}
    \right)  + \gamma\right]
  \Pc_{\nu\tau;\mu\eta}^{\left(j_{1};j_{2}\right)} \nonumber\\
  & = i\frac{\xi}{\mathcal{D}}
  f_{\nu}^{(j_1)}f_{\mu}^{(j_2)}\delta_{\nu\tau}
  \mathcal{C}_{\mu,\eta}^{\left(\sigma\right)}
  \pol_{\sigma}^{\ast} \cdot \Dv_{F}^{+}\left(\stochxv_{j_{2}}\right)  \nonumber\\
  & +i\xi
  \mathcal{C}_{\beta,\eta}^{\left(\sigma\right)}
  \pol_{\sigma}^{\ast} \cdot
  \radKernel^{\prime}\left(\stochxv_{j_{2}}-\stochxv_{j_{1}}\right)
  \pol_{\varsigma}
  \mathcal{C}_{\tau,\zeta}^{\left(\varsigma\right)}
  \Pc_{\mu\beta;\nu\zeta}^{\left(j_{2};j_{1}\right)} \nonumber\\
  & + i\xi \sum_{j_{3} \notin \{ j_{1},j_{2}\}}
  \mathcal{C}_{\beta,\eta}^{\left(\sigma\right)}
  \pol_{\sigma}^{\ast} \cdot
  \radKernel^{\prime}\left(\stochxv_{j_{2}}-\stochxv_{j_{3}}\right)
  \pol_{\varsigma}
  \mathcal{C}_{\epsilon,\zeta}^{\left(\varsigma\right)}
  \Pc_{\nu\tau,\mu\beta;\epsilon\zeta}^{\left(j_{1},j_{2};j_{3}\right)}\,,
  \label{eq:multi_atom_2_body_eqm}
\end{align}
where repeated indices $\beta$, $\epsilon$, $\zeta$, $\sigma$ and $\varsigma$
are summed over.
The first term on the right-hand-side of Eq.~\eqref{eq:multi_atom_2_body_eqm} represents
driving of the two-atom internal level correlation function by the incident field.
The second term on the right-hand-side is responsible for the recurrent scattering processes where an atomic polarization excitation
is repeatedly exchanged between atoms $j_1$ and $j_2$ that are fixed at the positions $\stochxv_{j_{1}}$ and $\stochxv_{j_{2}}$, respectively.
In the second term we can identify two types of dynamics: (i) recurrent scattering events for which $\nu=\tau$ and $\mu=\beta$, indicating that an excitation is exchanged between a ground-state atom and a polarization; (ii) processes in which one or both of the following are true: $\nu\neq\tau$ and $\mu\neq\beta$.

The case (i) represents classical electrodynamics of coupled dipoles, and in App.~\ref{app:multilevel} we show that these processes are already incorporated in  the multilevel classical approximation of Eq.~\eqref{eq:multi_level_P2_simple}.
The case (ii) describes a virtual scattering process between atoms $j_1$ and $j_2$.
This virtual scattering term is no longer classical because, even when the
atoms are initially in an incoherent superposition of Zeeman
levels, 
the
photon exchange process can generate correlations arising from virtual
fluctuations between Zeeman states in one atom and the dipole
amplitude of another.  Roughly speaking, as the dipole excitation
is transferred from one atom to another, the virtual emission and
reabsorption process can leave behind fluctuations between Zeeman
levels in the emitting atom that are correlated with the absorbing
atom's dipole amplitude.
We emphasize, however, that \emph{single}-atom ground state populations and
coherences, themselves, are unaffected
by virtual photon exchanges to first order in the electric field amplitude.

The last term of Eq.~\eqref{eq:multi_atom_2_body_eqm} is the
driving of a two-atom internal level correlation function by three-atom internal
level correlation functions that
represent the dipole
amplitude of an atom $j_3\ne j_1,j_2$ and atoms $j_1$ and $j_2$
being in the ground level.
In principle, to solve the optical response of the cloud exactly,
one would need to find the dynamics of the three-atom internal level correlation functions.
This would, in turn, require one to solve the dynamics of
four-atom internal level correlation functions, and so on.  The result is a hierarchy of
equations for the internal states of discrete emitters at fixed
positions.  This hierarchy is reminiscent of the full
density-polarization correlation function hierarchy presented in
App.~\ref{sec:monte-carlo-eval}~\cite{Ruostekoski1997a} when the atoms
are at fixed positions $\{\stochxv_1,\ldots,\stochxv_N\}$.

In App.~\ref{app:multilevel}  we show that if we only consider the case (i) of Eq.~\eqref{eq:multi_atom_2_body_eqm}, for which $\nu=\tau$ and $\mu=\beta$, and the corresponding hierarchy of equations for the internal level correlation functions with only the analogous diagonal terms $\Pc_{\nu_1\nu_1,\ldots,\nu_{\ell-1}\nu_{\ell-1};\nu_\ell\zeta}^{(j_1,\ldots,j_{\ell-1};j_\ell)}$ included, we can solve the entire hierarchy by the classical electrodynamics simulations of the previous section. For the $\ell$th order correlation functions we substitute
\beq
\Pc_{\nu_1\nu_1,\ldots,\nu_{\ell-1}\nu_{\ell-1};\nu_\ell\zeta}^{(j_1,\ldots,j_{\ell-1};j_\ell)} = f_{\nu_1}^{(j_1)}\ldots f_{\nu_\ell}^{(j_\ell)}\Pc_{\nu_\ell\zeta}^{(j_\ell)}\,.
\label{newcor1}
\eeq
With this substitution, the entire hierarchy can be reduced to the coupled time-dependent equations of multilevel dipoles [Eq.~\eqref{eq:Pcv_multi_level_dynamics} with Eq.~\eqref{eq:multi_level_P2_simple}]
\begin{align}
  {d\over dt} &\Pc_{\mu\eta}^{(j)} = (i\bar{\Delta}_{g\mu e\eta}-\gamma)\Pc_{\mu\eta}^{(j)}
  + i\frac{\xi}{\Dc}\,  \mathcal{C}_{\mu,\eta}^{(\sigma)} \pol_{\sigma}^\ast\cdot
  \Dv^+_F(\stochxv_j) \nonumber\\
  & +i\xi \sum_{l\neq j} {\cal C}_{\mu,\eta}^{(\sigma)}\pol_{\sigma}^\ast\cdot
  \radKernel'(\stochxv_j-\stochxv_l)  \pol_{\varsigma}\mathcal{C}_{\epsilon,\zeta}^{(\varsigma)}
 f_\epsilon^{(l)}\Pc_{\epsilon\zeta}^{(l)}\, \textrm{.}
  \label{eq:Pcv_multi_level_dynamics2a}
\end{align}

However, in order to solve the full quantum response we also need to include the terms of case (ii) of Eq.~\eqref{eq:multi_atom_2_body_eqm}.
To make such a solution tractable for a
system of many atoms, we can truncate the hierarchy at the level of
two-atom amplitudes.  To do this, we approximate
$\Pc_{\nu_{1}\tau_{1},\nu_{2}\tau_{2};\nu_{3}\eta_{3}}^{\left(j_{1},j_{2};j_{3}\right)}$
as a function of one and two atom quantities.
This function is not unique, and the optimal choice may depend on
the particular physical system.  For example one could
write, for an incoherent mixture
\begin{widetext}
  \begin{equation}
    \Pc_{\nu_{1}\tau_{1},\nu_{2}\tau_{2};\nu_{3}\eta_{3}}^{\left(j_{1},j_{2};j_{3}\right)}
    \approx f_{\nu_{1}}^{\left(j_{1}\right)}\delta_{\nu_{1}\tau_{1}}
    \Pc_{\nu_{2}\eta_{2};\nu_{3}\eta_{3}}^{\left(j_{2};j_{3}\right)}
    +f_{\nu_{2}}^{\left(j_{2}\right)}\delta_{\nu_{2}\tau_{2}}
    \mathcal{P}_{\nu_{1}\tau_{1};\nu_{3}\eta_{3}}^{\left(j_{1};j_{3}\right)}
    - f_{\nu_{1}}^{\left(j_{1}\right)}f_{\nu_{2}}^{\left(j_{2}\right)}\delta_{\nu_{1}\tau_{1}}\delta_{\nu_{2}\tau_{2}}
 	  \mathcal{P}_{\nu_{3}\eta_{3}}^{\left(j_{3}\right)} \,
          \textrm{.}
          \label{eq:three_body_trunc}
        \end{equation}
\end{widetext}
Substituting Eq.~\eqref{eq:three_body_trunc} into
Eq.~\eqref{eq:multi_atom_2_body_eqm}, one finds that
Eq.~\eqref{eq:multi_atom_2_body_eqm} forms a closed system of equations
for $\Pc_{\nu\tau;\mu\eta}^{\left(j_{1};j_{2}\right)}$.
The solution of these equations, in conjunction with
Eq.~\eqref{eq:Pcv_multi_level_dynamics},  gives the optical response for a
single realization of atomic positions
$\{\stochxv_1,\ldots,\stochxv_N\}$.
Ensemble averaging
over many realizations yields the optical response of the gas.

\section{Stochastic electrodynamics simulations of the optical response of
saturated atoms}

\label{sec:highintensity}

The full quantum field-theoretical representation for the cooperative
response of the ensemble of two-level atoms in the low light intensity limit can
be written as a hierarchy of $N$ equations of motion for correlation functions
-- with each correlation function involving only a single excited-state field operator  [given by the two-level
equivalent of Eq.~\eqref{eq:corrFuncEqsOfM}]. Outside of the low light intensity limit
this is no longer possible; an equivalent
calculation requires one to consider equations for all atomic correlation
functions.  Even in an ensemble of two-level atoms, one must track all $2^n$
$n$-body correlation functions for every $n<N$; multiple internal levels further
increase the complexity. Consequently, accounting for saturation makes the
solution of the optical response significantly more difficult.  Fortunately, the
classical electrodynamics simulation techniques used in the previous section can be extended
to include the effects of saturation. When the excited-state populations are not negligible we
explicitly keep track of the internal level dynamics of the atoms -- and introduce an approximation
that is therefore semiclassical in nature. We also incorporate the light-induced correlations
between the different atoms and different levels that depend on the spatial distribution of
the atoms, but ignore quantum-mechanical fluctuations between the different levels
[by means of factorizing many-body internal level correlation
functions for discrete position realizations, in a manner analogous to that of
Sec.~\ref{sec:lowintensitymultilevel}]. We
first discuss the simple case of two-level atoms before generalizing to the full
description of multilevel atoms.

\subsection{Two-level atoms}

\label{sec:HighIntensityTwoLevel}

As in the low light intensity case, we seek the solution of the equations
governing the optical response of two-level atoms in the presence of saturation
[Eqs.~\eqref{eq:dPdt_2levelhigh} and~\eqref{eq:drhogdt_2levelhigh}] by deriving
equations of motion for the electrodynamics of $N$ atoms located at a set of
discrete positions $\{\stochxv_1,\ldots,\stochxv_N\}$. The procedure removes the
spatial correlations in each individual stochastic realization and the
electrodynamics can be solved for a fixed set of atomic positions. The solution
to the optical response of an atomic ensemble with distributed atomic positions
is then obtained by means of averaging over many such realizations; this
restores the correct spatial correlations when the positions in each
realization are sampled from an appropriate $N$-body joint probability
distribution function for the system.

In order to derive the equations of motion for pointlike scatterers at fixed
discrete positions $\{\stochxv_1,\ldots,\stochxv_N\}$ for individual stochastic
realizations, we use a similar mathematical procedure as in the low light
intensity case of Sec.~\ref{sec:lowintensitytwolevel}. We take the expectation
values with respect to the center-of-mass coordinates of the atoms of both sides
of Eqs.~(\ref{eq:dPdt_2levelhigh}) and (\ref{eq:drhogdt_2levelhigh}) conditioned
on the $N$ atoms being at the positions $\stochxv_1,\ldots,\stochxv_N$, as if
the atoms were in a hypothetical physical system pinned at fixed positions.
However, since here we incorporate the saturation of the atoms, there also
exists additional internal-level dynamics beyond the spatial positions of the
atoms that can become correlated by the scattered light. The system therefore no
longer resembles the classical electrodynamics of a set of linear harmonic
oscillators and the internal states of different atoms can in
principle be nonclassically correlated. The conditioned expectation values of
one-body operators can be expressed in terms of one-body internal-level density matrix elements
\footnote{We use here the less common notation that $\rho_{ge}$
corresponds to the matrix element $\bra{e}\hat{\rho}\ket{g}$ of the density operator
$\hat{\rho}$.  The order of the subscripts is chosen to mirror the second-quantized
formalism used in the remainder of the paper, such that $\rho_{ge}$ is related to
$\av{\cpsi_g\dpsi_e}$.}
\begin{align}
\av{\cpsi_e\dpsi_e}_{\{\stochxv_1,\ldots,\stochxv_N\}} &= \sum_j
\rho_{ee}^{(j)}\delta(\rv-\stochxv_j), \label{eq:twolevelrhoedef}\\
\av{\cpsi_g\dpsi_g}_{\{\stochxv_1,\ldots,\stochxv_N\}} &= \sum_j
\rho_{gg}^{(j)}\delta(\rv-\stochxv_j),\label{eq:twolevelrhogdef} \\
\av{\Pvhat^{+}(\rv)}_{\{\stochxv_1,\ldots,\stochxv_N\}} &= \Dc\pol\sum_j
\rho_{ge}^{(j)}\delta(\rv-\stochxv_j)\,,\label{eq:sattwolevelPplusdef}
\end{align}
where we then require $\rho_{ee}^{(j)} +\rho_{gg}^{(j)} = 1$ for the
conservation of the atom population. In Eq.~\eqref{eq:sattwolevelPplusdef} we
have explicitly written the fixed orientation $\pol$ of the atomic dipoles for
two-level atoms.  The two-body density-polarization correlations from
Eqs.~(\ref{eq:dPdt_2levelhigh}) and (\ref{eq:drhogdt_2levelhigh}) for a single
realization can similarly be represented by two-body internal-level density matrix elements
\begin{align}
\av{\cpsi_a(\rv)&\cpsi_b(\rv')\dpsi_c(\rv')\dpsi_d(\rv)}_{\{\stochxv_1,\ldots,\stochxv_N\}}\nonumber
\\ &= \sum_{\substack{j,l=1 \\ j \neq l}}^N
\rho_{ad;bc}^{(j,l)}\delta(\rv-\stochxv_j)\delta(\rv'-\stochxv_l)\, .
\label{eq:saturatedtwolevelpaircorrelationdefn}
\end{align}
The matrix elements $\rho_{ad;bc}^{(j,l)}$ ($a,b,c,d=g,e$) represent internal level
pair correlations between atoms at the fixed points $\stochxv_j$ and $\stochxv_l$.
While, in the limit of low light intensity, only correlations involving at most one
excited state operator were non-negligible, at arbitrary intensities we must include
all 16 possible internal level pair correlations.  Coupled equations of
motion for such pair correlations can be obtained, following the method we earlier
used to obtain
Eq.~\eqref{eq:multi_atom_2_body_eqm}, which depend in turn upon three-body internal
level correlations, and so on. To develop a computationally efficient numerical
method, we introduce a semiclassical approximation for the internal level pair correlations in a single realization of discrete
atomic positions by factorizing them between different atoms,
analogously to the approximation introduced
in Sec.~\ref{sec:lowintensitymultilevel}.  For example, we take
\beq
\rho_{gg;ge}^{(j,l)} \simeq \rho_{gg}^{(j)}\rho_{ge}^{(l)}.
\label{eq:P2decorrelation}
\eeq

In fact, we made an analogous substitution earlier, in
Sec.~\ref{sec:lowintensitytwolevel}, for two-level atoms in the limit of low light
intensity (or indeed, any atoms with a single electronic ground level).  In that
limit, it was shown in Ref.~\cite{Javanainen1999a} and in
App.~\ref{sec:monte-carlo-eval} that the subsequent averaging over an ensemble of
many such discrete positions realizations reproduces the dynamics of the full
quantum field-theoretical representation of the correlation functions that govern
the optical response.  Within statistical accuracy, the stochastic electrodynamics
simulations therefore give the exact optical response for an ensemble of
stationary two-level atoms in this limit. However, the argument of
App.~\ref{sec:monte-carlo-eval} relies on the linearity of the the model of coupled
linear oscillators that results for each realization.

In contrast, in the saturated case the dynamics inherently involves the excited
level and therefore constitutes a multilevel system.
This semiclassical approximation stems from an analogous substitution to that
which led to the classical approximation of the pair correlations for atoms with
multiple ground levels in the low intensity case
[Sec.~\ref{sec:lowintensitymultilevel}].
Analogously to the
multilevel low light intensity case,
the scattered light may induce nonclassical correlations between the internal
levels of different atoms that no longer can be captured by classical stochastic
sampling and the electrodynamics of one-body operators, such as $\rho_{ge}^{(j)}$, by the factorization approximation
Eq.~\eqref{eq:P2decorrelation}. Specifically in this
case, the nonlinear saturation at higher intensities prevents a similar solution
to that of the low light intensity two-level case, and indeed the factorization
of two-body internal level correlations
introduces a semiclassical approximation.\footnote{Note, however, that we refer to the approximation here as
semiclassical, since we still include nonlinear effects due to saturation in a two-level system.}
In essence, we neglect internal state
correlations between atoms in a single realization for quantities similar to
$\av{\Pvhat^+(\rv)\Pvhat^-(\rv')}$, and averaging over many realizations does
not fully restore all the correlations. Terms like
$\av{\Pvhat^+(\rv)\Pvhat^-(\rv')}$ can exhibit nonclassical correlations between
the atoms that are excluded in our stochastic treatment. Nonclassical saturation
phenomena in analogous correlation functions are familiar from inelastic
scattering in the resonance fluorescence of a single atom and, e.g., in the
formation of the Mollow triplet~\cite{Mollow1969a}. In principle, one might
improve on this semiclassical approximation by accounting for nonclassical pair
correlations by considering the equations of motion for $\rho_{ad;bc}^{(j,l)}$
in a manner akin to that described in Sec.~\ref{sec:incorp-two-body}.

Under this factorization approximation, the expectation values of
Eqs.~(\ref{eq:dPdt_2levelhigh}) and (\ref{eq:drhogdt_2levelhigh}) for a given
set of fixed positions $\stochxv_1,\ldots,\stochxv_N$ of point scatterers then
lead to
\begin{align}
{d\over dt}\,&\rho_{ge}^{(j)}=(i\bar{\Delta}-\gamma)\rho_{ge}^{(j)} - i\frac{\xi}{\Dc}
\left(2\rho_{ee}^{(j)}-1\right) \pol^\ast\cdot\Dv_F^+(\stochxv_j) \nonumber\\
&-i\xi\left(2\rho_{ee}^{(j)}-1\right)\sum_{l\neq j} \pol^\ast\cdot {\sf
G}'(\stochxv_j-\stochxv_l)\pol \rho_{ge}^{(l)}, \label{eq:2levelsaturateddPdtstochastic}\\
{d\over dt}\,&\rho_{ee}^{(j)}= -2\gamma\rho_{ee}^{(j)}
+\frac{2\xi}{\Dc}\mathrm{Im}\left[\pol\cdot\Dv_F^-(\stochxv_j)\rho_{ge}^{(j)}\right] \nonumber \\
&+2\xi\mathrm{Im}\left[\sum_{l\neq j}
\rho_{ge}^{(j)}\pol\cdot{\sf G}'^*(\stochxv_j-\stochxv_l)
\pol^\ast\rho_{eg}^{(l)}\right],\label{eq:2levelsaturateddrhoetstochastic}
\end{align}
where we have made use of $\rho_{ee}^{(j)} +\rho_{gg}^{(j)} = 1$ to eliminate
$\rho_{gg}^{(j)}$ from Eqs.~\eqref{eq:2levelsaturateddPdtstochastic}
and~\eqref{eq:2levelsaturateddrhoetstochastic}.  Once again, we have assumed the
driving field to be in a coherent state, and so under expectation values the
operators $\Dvhat_F^\pm$ give rise to multiplicative classical coherent fields
$\Dv_F^\pm$.  This set of $2N$ nonlinear equations can then be solved for the
optical response of a single realization of point scatterers.  As in the low
light intensity limit, the equations have an intuitive form. For a pointlike
two-level atom at position $\stochxv_j$, with ground and excited state
amplitudes $\rho_{gg}^{(j)}$ and $\rho_{ee}^{(j)}$, respectively,
Eq.~\eqref{eq:2levelsaturateddPdtstochastic} gives the optical response to the
sum of the driving field $\Dv_F^+$ and fields scattered from each of the other
atoms $\Dc{\sf G}'(\stochxv_i-\stochxv_j)\pol\rho_{ge}^{(l)}$.  The stochastic
simulation of the total response of the system proceeds as in
Sec.~\ref{sec:lowintensitytwolevel}.  A set of atomic positions
$\{\stochxv_1,\ldots,\stochxv_N\}$ is chosen randomly from the $N$-body
probability distribution function, which we assume to be known.  The above
closed set of nonlinear equations can then be solved numerically, and ensemble
averaging over many such realizations generates quantities
$\av{\cpsi_e(\rv)\dpsi_e(\rv)}$, $\av{\cpsi_g(\rv)\dpsi_g(\rv)}$, and
$\av{\Pvhat^+(\rv)}$, detailing  the cooperative response of the system.  In
addition to the $N$-body probability distribution function, the nonlinear nature
of the system means that the initial values of $\rho_{ee}^{(j)}$ as well as
$\rho_{ge}^{(j)}$ are required, although in the common experimental situation of
response to a probe pulse the initial state in the absence of a driving field is
trivially $\rho_{ge}^{(j)} = \rho_{ee}^{(j)} = 0$.

\subsection{Multilevel atoms}
\label{sec:HighIntensityMultiLevel}

Realistic experimental situations frequently involve several internal atomic levels that
participate in optical probing. In the low light
intensity case we generalized the two-level system for the case of multilevel
atoms (Sec.~\ref{sec:lowintensitymultilevel}). In the saturated case we can
proceed in the same way and generalize the saturated two-level case of the
previous section to a multilevel formalism.   In contrast to the low intensity
case of Sec.~\ref{sec:lowintensitymultilevel}, at higher intensities the light
may drive dynamics between the relative populations of different internal
levels, complicating the analysis. Generalizing the earlier decomposition for a single realization of
point scatterers, and to account for the dynamics of level populations, we write
\beq
\av{\cpsi_{a\nu} \dpsi_{b\eta}}_{\{\stochxv_1,\ldots,\stochxv_N\}} =
\sum_j \rho_{a\nu b\eta}^{(j)}\delta(\rv-\stochxv_j)\,,
\eeq
where $a,b=e,g$.
Taking expectation values of Eqs.~(\ref{eq:fullresponseequations}) for a given
realization of discrete atomic positions now leads to
the equations for the response
\begin{subequations}
\begin{align}
{d\over dt}&\,\rho_{g\nu e\eta}^{(j)} =
(i\bar{\Delta}_{g\nu e\eta}-\gamma)\rho_{g\nu e\eta}^{(j)}\nonumber\\
&+ i\frac{\xi}{\Dc^2} \rho_{g\nu g\tau}^{(j)} \dv_{e\eta g\tau}
\cdot {\bf D}^{+}_F(\stochxv_j)\nonumber\\
&
- i\frac{\xi}{\Dc^2}
\rho_{e\tau e\eta}^{(j)}\dv_{e\tau g\nu}\cdot
{\bf D}^{+}_F(\stochxv_j) \nonumber \\
&+ i\frac{\xi}{\Dc^2}
\sum_{l\neq j}\,\dv_{e\eta g\tau}\cdot
{\sf G}'({\stochxv_j}-{\stochxv_l})
\dv_{g\mu e\zeta}\rho_{g\mu e\zeta}^{(l)}
\rho_{g\nu g\tau}^{(j)}\nonumber\\
& -i\frac{\xi}{\Dc^2}
\sum_{l\neq j}\,\dv_{e\tau g\nu}\cdot
{\sf G}'({\stochxv_j}-{\stochxv_l})
\dv_{g\mu e\zeta}\rho_{g\mu e\zeta}^{(l)}
\rho_{e\tau e\eta}^{(j)}
\label{Pfullstochastic}
\,,
\end{align}
\begin{align}
{d\over dt}\,&\rho_{g\nu g\eta}^{(j)} =
i\bar{\Delta}_{g\nu g\eta}
\rho_{g\nu g\eta}^{(j)}
+2\gamma {\cal C}_{\eta,(\eta+\sigma)}^{(\sigma)} {\cal C}_{\nu,(\nu+\sigma)}^{(\sigma)}
\rho_{e(\nu+\sigma) e(\eta+\sigma)}^{(j)}
\nonumber \\
&-i\frac{\xi}{\Dc^2}\rho_{e\tau g\eta}^{(j)}
\dv_{e\tau g\nu}\cdot
{\bf D}^{+}_F(\stochxv_j)\nonumber \\
&
+i\frac{\xi}{\Dc^2}\rho_{g\nu e\tau}^{(j)}\dv_{g\eta
e\tau}\cdot{\bf D}^{-}_F(\stochxv_j) \nonumber \\
&-i\frac{\xi}{\Dc^2}\sum_{l\neq j}\,\dv_{e\tau g\nu}\cdot{\sf G}'({\stochxv_j}-{\stochxv_l})
\dv_{g\mu e\zeta}\rho_{g\mu e\zeta}^{(l)}
\rho_{e\tau g\eta}^{(j)} \nonumber\\
&+i\frac{\xi}{\Dc^2}\sum_{l\neq j}\,\dv_{g\eta e\tau}\cdot{\sf G}'^*({\stochxv_j}-{\stochxv_l})
\dv_{e\zeta g\mu}\rho_{e\zeta g\mu}^{(l)}
\rho_{g\nu e\tau}^{(j)}\,,
\label{eq:drhog_fullstochastic}
\end{align}
\begin{align}
{d\over dt}\,&\rho_{e\nu e\eta}^{(j)} =
(i\bar{\Delta}_{e\nu e\eta}
-2\gamma)\rho_{e\nu e\eta}^{(j)}
\nonumber \\
&+i\frac{\xi}{\Dc^2}\rho_{e\nu g\tau}^{(j)}\dv_{e\eta g\tau}\cdot
{\bf D}^{+}_F(\stochxv_j)\nonumber
\\
&-i\frac{\xi}{\Dc^2}\rho_{g\tau e\eta}^{(j)}
\dv_{g\tau e\nu}\cdot{\bf D}^{-}_F(\stochxv_j) \nonumber \\
&-i\frac{\xi}{\Dc^2}\sum_{l\neq j}\,
\dv_{g\tau e\nu}\cdot{\sf G}'^*({\stochxv_j}-{\stochxv_l})
\dv_{e\zeta g\mu}\rho_{e\zeta g\mu}^{(l)}
\rho_{g\tau e\eta}^{(j)} \nonumber \\
&+i\frac{\xi}{\Dc^2}\sum_{l\neq j}\, \dv_{e\eta g\tau}\cdot
{\sf G}'({\stochxv_j}-{\stochxv_l})
\dv_{g\mu e\zeta}\rho_{g\mu e\zeta}^{(l)}
\rho_{e\nu g\tau}^{(j)}\,
\label{eq:drhoe_fullstochastic},
\end{align}
\label{eq:multilevelhighintensitystochastic}
\end{subequations}
where the repeated indices $\tau,\zeta,\mu,\sigma$ are implicitly summed over,
and $\bar{\Delta}_{a\nu b\eta} = \Delta_{b\eta}-\Delta_{a\nu}$.
In the case of a conserved total atom population, one of the equations can be eliminated by the relation
$ \sum_\nu \rho_{g\nu g\nu}^{(j)} +\sum_\eta \rho_{e\eta e\eta}^{(j)}=1$.
As in the
two-level case, we have  introduced a semiclassical approximation to factorize
internal level two-body correlation functions  in the manner of
Eq.~\eqref{eq:P2decorrelation}. Due to this approximation, as discussed in the
two-level case of the previous section, the ensemble average of many single
realizations does not reproduce the nonclassical correlations in the system.
To facilitate comparison with the low light intensity results of
Sec.~\ref{sec:lowintensitymultilevel}, we note that the induced electric
dipole of atom $j$ for the optical transition
$|g,\nu\rangle\rightarrow |e,\eta\rangle$ is
$\dv_{g\nu e\eta}\rho_{g\nu e\eta}^{(j)}$, which corresponds to the quantity $\Dc\pol_{\eta-\nu}
\mathcal{C}_{\nu,\eta}^{(\eta-\nu)} f_\nu^{(j)} \Pc_{\nu\eta}^{(j)}$ in the
notation of Sec.~\ref{sec:lowintensitymultilevel}.

The result is a set of coupled
equations~\eqref{eq:multilevelhighintensitystochastic} for internal level
one-body density matrix elements $\rho_{a\nu b\eta}^{(j)}$, for each atom
$j=1,\ldots,N$.  We stochastically sample the set of discrete atomic coordinates
$\{\stochxv_1,\ldots,\stochxv_N\}$ from the $N$-body density distribution
function. For each realization we then solve the semiclassical electrodynamics equations
of motion~\eqref{eq:multilevelhighintensitystochastic} for $\rho_{a\nu b\eta}^{(j)}$.
Averaging over many such realizations allows the expectation
values of desired observables to be computed.
The nonlinear nature of the equations means that the initial values of
$\rho_{a\nu b\eta}^{(j)}$ are required that may, e.g., include nonvanishing initial coherences.

\section{Numerical example -- Atoms in a periodic 2D lattice}
\label{sec:mott-insulator-state}

In this section, we provide an example of the stochastic simulation techniques described in
this paper by studying a Mott-insulator state of atoms confined in a 2D optical lattice
that is subjected to incident low-light-intensity excitation.
This system has previously been considered in Ref.~\cite{Jenkins2012a}. We analyze here in detail the phenomena
that are most closely related to the present work. We also extend the results of Ref.~\cite{Jenkins2012a} to calculate the collective
excitation eigenmode distributions of the atoms.

We show that the atoms even in a 2D lattice geometry can respond cooperatively
when the scattered light mediates strong dipole-dipole interactions between atoms.
We find that the
optical excitations of the atoms to resonant light are collective and cannot be described by single atom
excitation properties.
The optical response is then governed by the distinct responses of the various collective modes to the incident
field rather than simply the sum of independent atoms.
The most subradiant of the collective excitation eigenmodes exhibits radiative resonance linewidths that are
dramatically narrowed.
Furthermore,
the narrowing exhibits a strong dependence on the light-induced interaction strength
between the atoms and is more sensitive to the spatial separation of the lattice sites
than to the lattice site confinement.
The optical response example for a phase-modulated incident field demonstrates how
light-induced correlation effects can be employed for engineering optical excitations on a subwavelength scale~\cite{Jenkins2012a}.
These correspond to electromagnetic energy hot-spots whose
widths beat the diffraction limit in an analogy to similar examples in solid-state circuit
resonator systems~\cite{KAO10}.

\subsection{Mott-insulator state of atoms in a lattice}

One of the central experimental developments in characterizing strongly
interacting phases of ultracold atoms in optical lattices is the ability to
manipulate and image atomic spins in individual lattice sites by
lasers~\cite{BakrEtAlNature2009,ShersonEtAlNature2010,singlespin}. In this
section, we use the classical electrodynamics simulations of point dipoles of
Sec.~\ref{sec:lowintensitytwolevel} to study the optical response of
a Mott-insulator state of atoms confined in a 2D optical lattice. For simplicity,
we consider two-level atoms when there is precisely one atom per lattice site.
Single-atom site-occupancy can be realized, for
example, by cooling bosonic atoms to the typical `wedding-cake' Mott-insulator
ground state of an optical lattice superposed on a weak harmonic trap, and
manipulating the sites with excess occupancy \cite{singlespin}.

We consider an $N_x \times N_y$ 2D square optical lattice with lattice
spacing $a$ and sites labeled
by index $i = 1,\ldots, N$ ($N=N_xN_y$), centered on positions $\Rv_i$
in the $xy$ plane.
Each site has a potential depth $sE_R$, where $E_R =
\pi^2\hbar^2/(2ma^2)$ is the lattice-photon recoil energy
\cite{Morsch06}.
The lattice sites are sufficiently deep that the optical lattice
potential around $\Rv_i$ is well approximated by a harmonic oscillator
with vibrational frequency $\omega = 2\sqrt{s}E_R/\hbar$.
Each occupied site $i$ contains a single atom in the vibrational
ground state which is described by the Wannier function
$\phi_i(\rv) \equiv \phi(\rv -
\Rv_i)$, where $\phi(\rv)$ is approximately the ground state of the
harmonic oscillator potential
\begin{equation}
  \label{eq:Wanier1}
  \phi(\rv) \simeq \frac{1}{(\pi^3\ell_x^2\ell_y^2\ell_z^2)^{1/4}}
  \exp\left(-\frac{x^2}{2\ell_x^2}-\frac{y^2}{2\ell_y^2} -
  \frac{z^2}{2\ell_z^2}\right) \text{.}
\end{equation}
The widths of the wavefunction in the lattice are $\ell_x=\ell_y=\ell =
as^{-1/4} / \pi$,  and we have assumed an additional oblate external
potential confines the atoms in a region around the $z=0$ plane with
thickness $\ell_z$.
For simplicity in the following examples, we assume $\ell_z \ll \ell$,
and neglect fluctuations of atomic position in the $z$ direction.
The atom density, $\rho_i(\rv) \equiv |\phi_i(\rv)|^2$, at site
$i$  thus has a Gaussian profile with a $1/e$ radius
$\ell$ in the $xy$ plane.
This radius is directly proportional to
the lattice spacing and narrows with increased trapping strength $s$.

\subsection{Light-atom interactions}

In the low light intensity limit, the
atoms behave as harmonic oscillators, as described in
Sec.~\ref{sec:lowintensity}.  Here we assume the atoms in each site have a
two-level structure with ground state $\ket{g}$, excited state $\ket{e}$, and
all other levels shifted out of resonance as described in
Sec.~\ref{sec:lowintensitytwolevel}. The orientation $\pol \approx \unitvec{e}_z
+ 0.1 \unitvec{e}_y$ of the atomic dipole transition is rotated slightly from
the normal to the lattice, and the dipole amplitude of atom $j$ is governed by
the oscillator variable $\Pc^{(j)}$. The fields emitted from each atom drive
oscillations in the others, and mediate interactions that produce the
cooperative optical response.  The dynamics are governed by
Eq.~\eqref{eq:Pol_single_atom_two_level}, which can be expressed in the form
\begin{equation}
  \label{eq:opt_lat_dynamics}
  \frac{d}{dt} \Pc^{(j)} = \sum_{l=1}^N \mathcal{M}_{j,l}\Pc^{(l)} +
  F^{(j)} \, \textrm{,}
\end{equation}
where the matrix $\mathcal{M}$ accounts for the single atom decay,
atomic level shifts, and the light-mediated interactions between different atoms.
The driving caused by the incident field is represented by $F^{(j)}$.
Each eigenvector of $\mathcal{M}$ corresponds to a collective mode
whose collective shift $\delta_c$ and decay rate $\Gamma$ are
determined by the mode's eigenvalue; $\Upsilon = - \Gamma -
i\delta_c$ \cite{JenkinsLongPRB,Jenkins2012a}.

\subsection{Sampling of atomic positions}

The stochastic realizations of atomic
positions $\{\stochxv_1,\stochxv_2,\ldots,\stochxv_{N}\}$ are sampled from the
joint probability distribution determined by the many-body wave function
$P(\rv_1,\ldots,\rv_N) = |\Psi (\rv_1,\ldots,\rv_N)|^2$ [see
Eq.~\eqref{PDT}].
We consider the limit where the mode functions for distinct
lattice sites $i_1$ and $i_2$ have negligible overlap, i.e.,
$\phi_{i_1}(\rv) \phi_{i_2}(\rv) \simeq 0$ for all $\rv$.
In this limit, the absolute square of the symmetrized many-body bosonic wave function is
\beq
  |\Psi (\rv_1,\ldots,\rv_N)|^2 \simeq {1\over N!}\sum_{i_1\ldots i_N}
   |\phi_{i_1} (\rv_1) |^2\ldots |\phi_{i_N} (\rv_N) |^2 \, \textrm{.}
  \label{manybody}
\eeq
where the summation runs over $N$-tuples $(i_1,\ldots,i_N)$ corresponding to all
permutations of sites $i=1,\ldots,{}N$ containing an atom.
Rather than sampling the positions of atoms $\rv_j$ labeled by index $j$
(which could be in any occupied site $i$), one can equivalently sample the
positions $\bar\rv_i$ of the atom in occupied site $i$ (which could
correspond to an atom labeled by any index $j$) from
the joint probability distribution
$\bar{P}(\bar\rv_1,\ldots,\bar\rv_N)$.
Since the many-body correlation function $P$ is
invariant under
permutation of atomic positions, the joint
probability distribution for atomic positions within particular
lattice sites is
\begin{align}
  \bar{P}(\bar\rv_1,\ldots,\bar\rv_N) &= \sum_{i_1,\ldots,i_N}
  P(\bar\rv_{i_1},\ldots,\bar\rv_{i_N}) \nonumber \\
  & = |\phi_{1} (\bar\rv_1) |^2\ldots
  |\phi_{N} (\bar\rv_N)|^2\,.
  \label{eq:P_bar_opt_lat}
\end{align}
Here, again, the $N$-tuples
$(i_1,\ldots,i_N)$ are summed over all permutations of sites
$i=1,\ldots,N$,
and we have taken advantage of the fact that, since atoms are
well-localized within their respective sites,
$|\phi_{i_1}(\bar\rv_{i_2})|^2$ is only nonzero  when $i_1=i_2$.
For each stochastic realization, we can thus treat the positions of
atoms within each site as independent Gaussian random variables with
mean $\Rv_i$ and variance $\ell^2/2$ in the $xy$ plane.

\subsection{Collective excitation eigenmodes}

\begin{figure}
  \centering
  \includegraphics[width=0.5108\columnwidth]{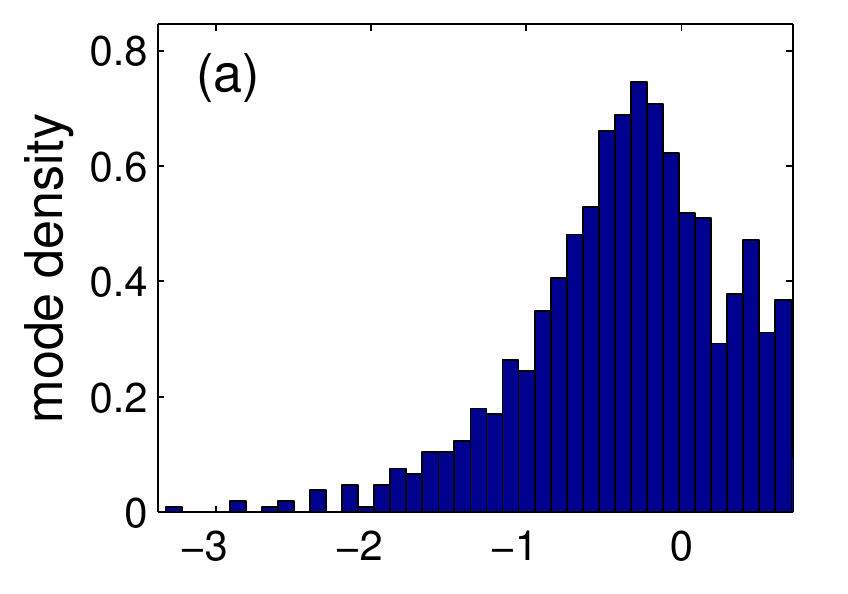}
  \includegraphics[width=0.4692\columnwidth]{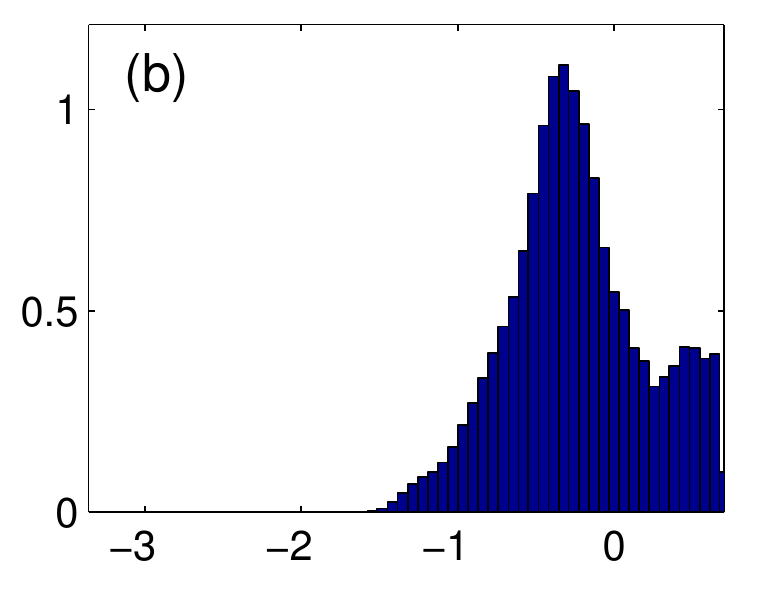}
  \includegraphics[width=0.5108\columnwidth]{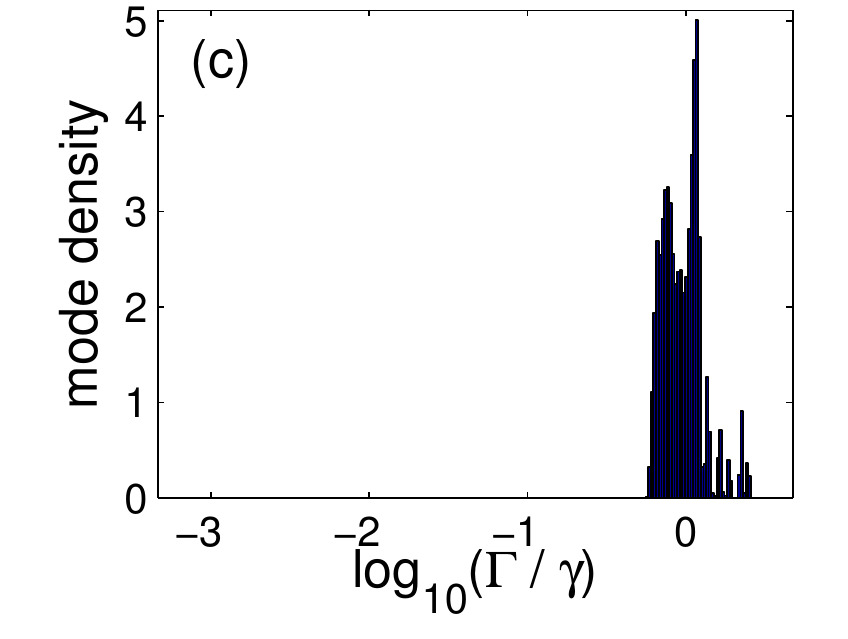}
  \includegraphics[width=0.4692\columnwidth]{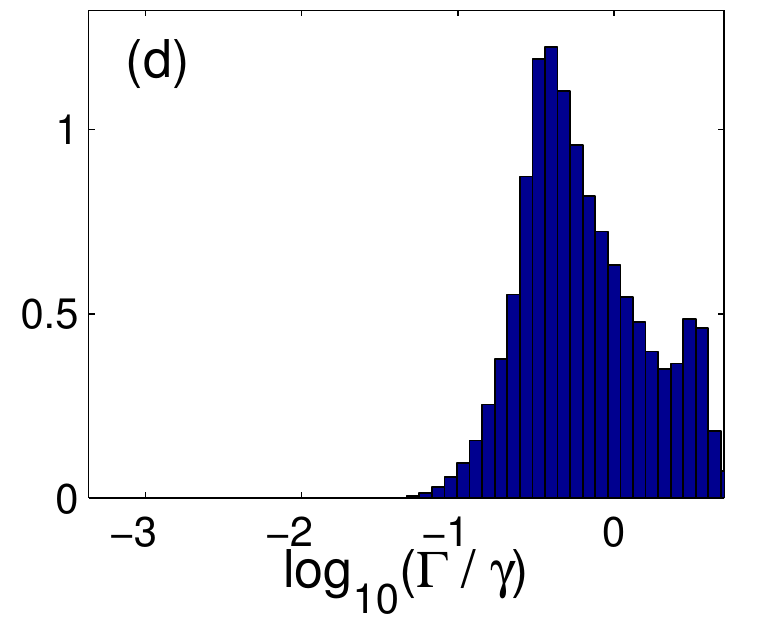}
  \caption{The probability distribution (normalized to one) of the logarithm of collective decay rates
    for two-level atoms in an optical lattice with dipole orientations
    $\pol \approx \unitvec{e}_z + 0.1 \unitvec{e}_y$. The lattice
    spacing is $a=0.55\lambda$ (a,b,d), while the lattice is only
    partially populated for panel c with an effective lattice spacing
    $a'=3a$. The widths of the Wannier functions are $\ell=0$ (a),
    $\ell=0.12a$ (b,c) and $\ell=0.21a$ (d). }
  \label{fig:mode_decay_dens}
\end{figure}

To characterize the cooperative properties of the system, we calculate its
collective excitation eigenmodes for many thousands of stochastic realizations of
atomic positions. Each eigenmode is associated with a collective decay rate and we obtain a
distribution of collective decay rates supported by the optical lattice.
Specifically, if one were to choose a realization of atomic positions
according to Eq.~\eqref{eq:P_bar_opt_lat} and, given those positions,
randomly select a collective mode of excitation with collective decay
rate $\Gamma$, the probability
density of obtaining a particular value of $\log_{10}(\Gamma/\gamma)$
corresponds to the densities shown in Fig.~\ref{fig:mode_decay_dens}.
Specifically, Figs.~\ref{fig:mode_decay_dens}(a,b,d)
show the distribution for a $32\times32$ optical
lattice with lattice spacing $a=0.55\lambda$.
When the lattice is infinitely deep
(Fig.~\ref{fig:mode_decay_dens}a), the atoms are perfectly
confined at the bottom of the lattice sites, and collective decay rates
range from the very subradiant $4.7\times 10^{-4}\gamma$ to the superradiant
$6.5\gamma$.
The fluctuations in atomic positions when $\ell\simeq 0.12 a$
(Fig.~\ref{fig:mode_decay_dens}b), associated with
lattice depth of $s=50$, narrow the
distribution of collective decay rates, which range between
$0.017\gamma$ and $6.3\gamma$.
The relatively shallow lattice with $s=5$ ($\ell\simeq 0.21 a$)
(Fig.~\ref{fig:mode_decay_dens}d) largely preserves the
distribution of decay rates calculated for a depth $s=50$.
Weakening the confinement strength from $s=50$ to $s=5$ reduces
the density of subradiant modes with decay rates below about
$0.1\gamma$.  However, the weakly confined lattice still permits
strongly subradiant modes.
For the realizations of atomic positions we sampled in the shallow ($s=5$)
lattice, decay rates range from  $3.5\times 10^{-3}\gamma$ to
$5.9\gamma$.
The distributions spanning several orders of magnitude indicate
the atoms in the lattice can exhibit strong light-induced correlations.

On the other hand, reducing the density of the atoms has a more deleterious effect
on the width of the distribution of collective decay rates than
reducing the tightness of confinement.
Figure~\ref{fig:mode_decay_dens}c shows the distribution of
collective decay rates in a lattice in which only every third lattice
site along the $x$ or $y$ directions is occupied.
In effect, the lattice spacing is widened to $a'=3a=1.65\lambda$.
Where the lattice with subwavelength spacing exhibited subradiant
collective decay rates suppressed by more than 3 orders of magnitude,
the decay rates in the sparsely populated lattice range from
$0.55\gamma$ to $2.6\gamma$.
When the atom density drops, light-induced correlations are weakened.

\subsection{Response to incident light}

\begin{figure}
  \centering
  \includegraphics[width=0.80\columnwidth]{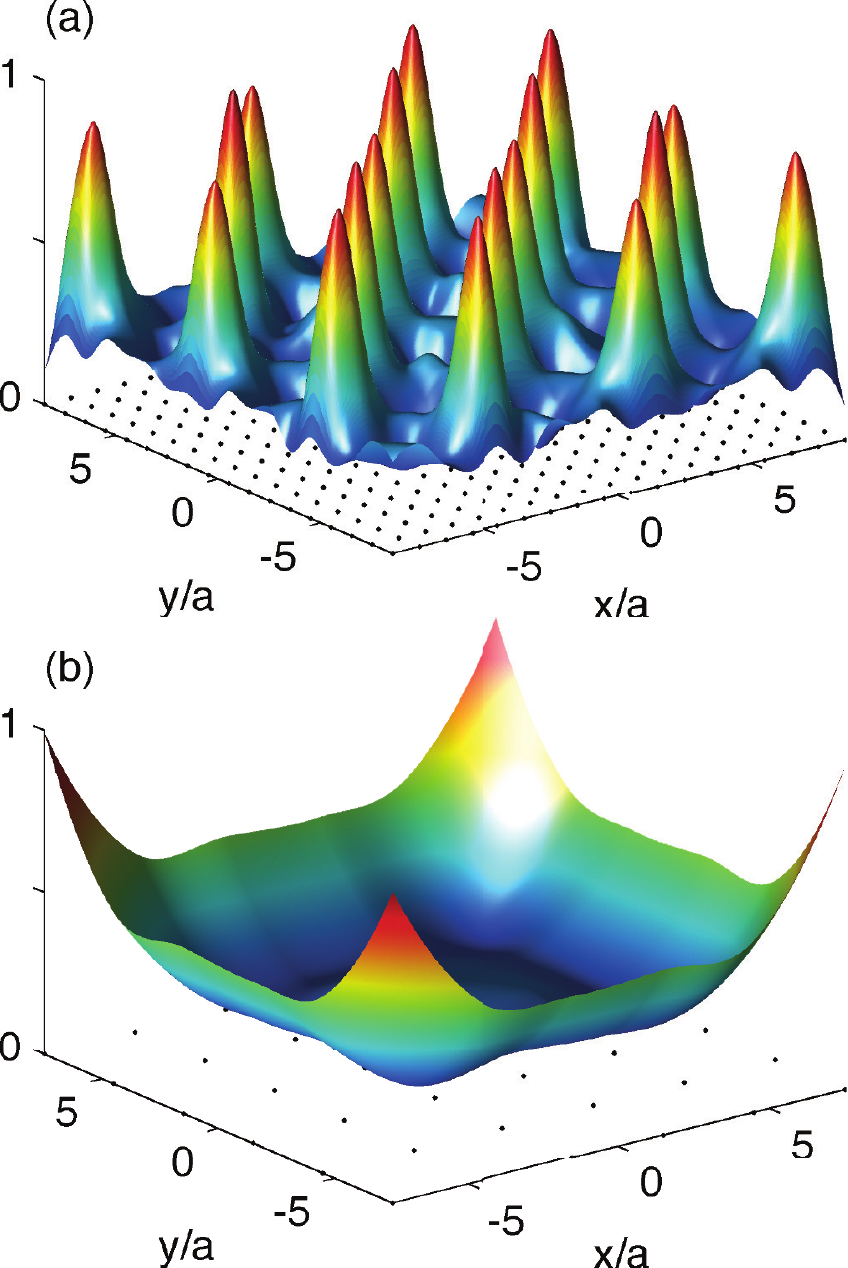}
  \caption{The atomic dipole intensities $\langle |\Pc^{(i)}|^2
    \rangle / \max_{i'}\langle |\Pc^{(i')}|^2\rangle$ in lattice
    sites $i$. Panel (a) shows strong light-induced correlations between the atoms in a lattice
    with spacing $a=0.55\lambda$ and depth $s=50$ to a phase-modulated
    driving [Eq.~\eqref{eq:phase_mod}] with the modulation period $ 6 a $.  Panel (b)
    illustrates the absence of cooperative response to a driving with modulation period $6 a'$ when only every third
    lattice site along each direction is occupied, giving an effective spacing
    of $a'=3a$.}
  \label{fig:coop_resp_lat}
\end{figure}

Next we compute
the response of the lattice to a phase-modulated incident field
\cite{Jenkins2012a}.
Specifically, we consider an $18 \times 18$ lattice ($a=0.55\lambda$)
with $s=50$.
For each realization of atomic positions, we compute the
atomic dipole amplitudes $\Pc^{(i)}$ in each site $i$
from the steady-state solution of  Eq.~\eqref{eq:opt_lat_dynamics}.
The external field drives each atom with equal
amplitude but with a phase $\varphi$ that varies with the $x$ and $y$
coordinates of the atomic positions as
\begin{equation}
  \varphi(x,y) \approx \frac{\pi}{2}
  \sin\left(2\pi \frac{x}{\Lambda a} \right) \sin\left(2\pi
    \frac{y}{\Lambda a}\right) \textrm{ .}
\label{eq:phase_mod}
\end{equation}
The incident field is only approximately represented by a field with
constant amplitude and sinusoidally modulated phase because such a
phase-modulation would involve evanescent waves -- a more accurate representation
may be obtained by a truncated superposition of plane waves as discussed
in Ref.~\cite{Jenkins2012a}.
Figure~\ref{fig:coop_resp_lat}a shows the
response of a system, in which each lattice site contains one atom, to such a field
tuned to the single-atom resonance. The modulation has a period of six lattice
sites ($\Lambda=6$). As a consequence of cooperative interactions, the phase
modulated driving excites a superposition of collective modes generating a
checkerboard pattern of the average excitation intensities $\langle |\Pc^{(i)}|^2
\rangle$ in which an atom at every sixth lattice site along each
direction is strongly excited. The widths (full width at half maximum) of the excitations are
less than the wavelength $0.9\lambda$.
This is significantly narrower than the period of the phase variation in the incident field, indicating that the localization length scale results from the interatomic interactions of the closely-spaced atoms.
When the effective lattice spacing is tripled to
$a'=3a$ as discussed above, however, the light-induced correlations are
weakened. A modulation with a comparable period of $6a'$ ($\Lambda=18$) results in uniform
excitation of sites inside the lattice, with atoms at the edge being strongly
excited due to boundary effects, as shown in
Fig.~\ref{fig:coop_resp_lat}.
The lattice spacing that produces a wide distribution of collective
decay rates (as in Fig.~\ref{fig:mode_decay_dens}b) results in strong light-induced correlations between the atoms in Fig.~\ref{fig:coop_resp_lat}a, while the lattice with a
narrow distribution (as in Fig.~\ref{fig:mode_decay_dens}b) does not
show similar behavior.

\section{Concluding remarks}

We have derived stochastic simulation techniques for the electrodynamics of atomic ensembles that scale favorably with atom number and are suitable also for dense atomic gases and
for light that is close to the atomic resonance frequency.
We first formulated quantum field-theoretical equations of motion describing the optical response [Eqs.~\eqref{eq:fullresponseequations}] that systematically include
the atomic saturation effects and the internal level structure.
In order to solve the resultant dynamics, we then introduced a simple procedure to derive stochastic electrodynamics equations
that can be used to study recurrent, or dependent, scattering between the atoms in the presence of strong light-atom coupling.
In the low light intensity
limit, stochastically sampling the positions of atoms from the
appropriate $N$-body density distribution function leads to a model of classical
coupled linear harmonic oscillators.  The solution of this model, and subsequent
averaging  over a large number of realizations, generates results for the optical
response of the atomic ensemble. The solution for coherently scattered light is
exact for stationary atoms possessing only a single electronic ground level,
since the averaged results reproduce the correct dynamics for the many-body
polarization density correlation functions to all orders, as shown in
App.~\ref{sec:monte-carlo-eval}.  We have extended this technique to deal with
atoms with multiple ground levels, and situations where saturation effects are
important.

In deriving our stochastic classical electrodynamics simulations, we have first
removed the spatial dependence from many-body correlation functions by considering
hypothetical cases of fixed atomic positions, sampling those positions from the
$N$-body joint probability distribution.  Averaging over many realizations later
restores the spatial aspect of the correlations, incorporating
recurrent scattering processes in the atomic response.   For a
single realization, the remaining many-body correlations are between internal levels.
To obtain numerically efficient simulations we have factorized such
internal-level correlations.  In general, this factorization can neglect
some many-body internal-level quantum correlations, and is therefore only
approximately valid.  The exception to this is the case of atoms with a single
ground level in the limit of low light intensity, where the stochastic
simulations give the exact response for the coherent scattering  of light from an
ensemble of stationary atoms.  Physically, this stems from the fact that there
are no dynamics of the excited populations (due to low intensity) and there are
no other ground levels involved in the optical transitions that could become
correlated. Therefore, for such a system the excitation of each atom can be
described by a linear polarisation amplitude, with no additional internal-level
dynamics possible.  The dynamics is equal to that of a coupled system of charged
classical linear harmonic oscillators and light, the system modeled by the
stochastic electrodynamics simulations in this limit
[Eq.~\eqref{eq:Pol_single_atom_J0_J1}].

In contrast, for atoms with multiple electronic ground levels, and/or in the presence
of saturation, the factorization of pair correlations provides a technique that
remains efficient and scales favorably with atom number, but at a cost of
neglecting some quantum internal-level many-body correlations.   We emphasize that in such situations, while some internal-level
pair correlations are not exactly reproduced, the ensemble averaging over many
stochastic realizations still incorporates recurrent scattering into the results of the stochastic simulations
(and, for example, for the case of the low light intensity limit multilevel system, the classical simulation approximation correctly predicts the expectation values of the different ground-state populations).  We have additionally
shown how one can improve on the approximate methods obtained by
this factorization by considering the dynamics of the internal-level pair
correlations, at a cost of a more complicated numerical system of
equations.

Light-induced correlation phenomena in wave propagation are strongly enhanced in
cold homogeneously broadened samples. Consequently, the stochastic simulations
techniques that incorporate recurrent scattering events between closely-spaced
atoms are most suitable in the systems where the inhomogeneous broadening (e.g.,
due to Doppler shifts) does not significantly exceed the radiative linewidth of
the atoms. In this limit the inhomogeneous broadening has been shown to suppress
the light-induced correlations~\cite{Javanainen2014a}, resulting in an optical
response that is more closely reminiscent of that of a standard continuous
medium electrodynamics.

The stochastic simulation methods replace the difficult problem of
deriving, and solving, many coupled equations of motion for up to $N$-body
correlations with a stochastic approach that relies on sampling
atomic positions. If the atoms are initially uncorrelated before the light enters
the medium, the position of each atom becomes an independent random variable
that is sampled from the atom density distribution.
In the presence of nontrivial position correlations for the atoms
the sampling from the correct many-body density operator for the system can
become challenging.   An ensemble of stochastic atomic positions must be
synthesized which generates the corresponding physical position correlations
between the atoms. While we have here investigated a relatively simple case of a
Mott-insulator state in an optical lattice
(Sec.~\ref{sec:mott-insulator-state}), systems with more complicated density correlation
functions, for example Fermi-Dirac statistics, can  also be sampled by Monte-Carlo simulation methods~\cite{Javanainen1999a}.

\acknowledgments

We acknowledge discussions with Antoine
Browaeys and Juha Javanainen.
This research was supported by the EPSRC and the Leverhulme trust.

\appendix

\section{Evaluation of the atom-light commutators}
\label{app:commutator}

Here we demonstrate the calculation of the commutators between the atomic field operators and the free electric displacement. Consider
$[\Dvhat_F^+({\bf r}),\hat\psi_{g\nu}({\bf r'})]\,$.
The total electric displacement $\Dvhat^+({\bf r})=\Dvhat^+_F({\bf r})+
\Dvhat^+_S({\bf r})$ is a canonical momentum of the light field and commutes
with the atomic field operators. Hence
\begin{align}
[\Dvhat_F^+({\bf r}),\hat\psi_{g\nu}({\bf r'})] =& -[\Dvhat_S^+({\bf r}),\hat\psi_{g\nu}({\bf r'})]\nonumber\\
= & -\int d^3 r' {\sf G}(\rv-\rv'') [\Pvhat^+(\rv''), \hat\psi_{g\nu}({\bf r'})]\nonumber\\& -[\Pvhat^+(\rv), \hat\psi_{g\nu}({\bf r'})]\,.
\end{align}
Applying then the standard commutation relations, such as
\beq
[\hat\psi_{g\tau}({\bf r}), \hat\psi^\dagger_{g\nu}({\bf r'})] =\delta(\rv-\rv') \delta_{\tau,\nu}\,,
\eeq
yields the result of Eq.~\eqref{eq:DFpsicomm}.

\section{Monte-Carlo evaluation of correlation functions in the low light intensity limit}
\label{sec:monte-carlo-eval}

By treating atoms and light as quantum fields throughout the analysis, it was
shown in Ref.~\cite{Ruostekoski1997a} that a cooperative coupling between atoms
and light leads to coupled equations of motion between correlation functions of
atomic density and polarization.  In the limit of low light intensity the
resulting hierarchy of equations represents strong light-mediated correlations
that become important at low temperatures and in dense samples--on the other
hand, for inhomogeneously broadened hot atoms the standard mean-field theory of
continuous effective medium electrodynamics can be
restored~\cite{Javanainen2014a}.  It was shown in Ref.~\cite{Javanainen1999a}
for a simplified 1D scalar electrodynamics how stochastically sampling atomic
positions, calculating the collective response for atoms at fixed positions, and
averaging over those stochastic realizations, reproduces the dynamics described
by these correlation functions. In this Appendix, we extend the quantum
field-theoretical representation that is expressed in terms of the hierarchy of
correlation functions~\cite{Ruostekoski1997a} to the general case of atoms with
multiple electronic ground levels. We then briefly generalize the treatment of
Ref.~\cite{Javanainen1999a} to the 3D electrodynamics including the full vector
properties of the electromagnetic fields for the special case of atoms with the
$J=0\rightarrow J'=1$ transition, corresponding to an isotropic susceptibility.
The analysis is similar to other atomic level systems that have only a single
electronic ground state, and shows how the coupled dynamics of correlation
functions in the presence of a single electronic ground level is reproduced by
the classical electrodynamics simulations.

We begin by considering the hierarchy of correlation
functions.
In the low light-intensity limit, the equation of motion for
polarization  simplifies to [Eq.~\eqref{eq:CR1}]
\begin{align}
  {d\over dt}\,\Pvhat^+_{\nu\eta} &=
  (i\bar{\Delta}_{g\nu e\eta}-\gamma)\Pvhat^+_{\nu\eta}
  + i\xi \hat\psi^\dagger_{g\nu}\hat\psi_{g\tau} {\sf P}^{\nu\eta}_{\eta\tau}
   \Dvhat^+_F \nonumber\\
  & + i\xi
  \int d^3r'\,{\sf P}^{\nu\eta}_{\eta\tau}
  {\sf G}'({\bf r}-{\bf r'})\,\hat\psi^\dagger_{g\nu}\Pvhat^+(\rv')
  \hat\psi_{g\tau}\,.
  \label{eq:CR1_repeated}
\end{align}
To determine the average scattered fields in the low light intensity limit,
it is necessary to find the average polarization within the atomic
ensemble.
Taking the expectation value of Eq.~\eqref{eq:CR1_repeated}, however, one
finds that the dynamics of the polarization depend, not simply on the
polarization density in the ensemble, but on a two-body correlation
function $\langle \hat\psi^\dagger_{g\nu}(\rv)\Pvhat^+(\rv')
  \hat\psi_{g\tau}(\rv)\rangle$. It was shown in
Ref.~\cite{Ruostekoski1997a} that in the low light
intensity limit, in order to solve the second-order correlation
function, one has to further find the dynamics of a third-order
correlation function; a third-order correlation function requires a
fourth-order correlation function, and so on.
The result is a hierarchy of dynamic equations of motion for the
$\ell^{\textrm{th}}$ order correlation functions

\begin{widetext}
  \begin{align}
    \label{eq:DensityCorrFuncDef_general}
    \rho_{\ell}^{\nu_{1}\tau_{1},\ldots,\nu_{\ell}\tau_{\ell}}\left(\rv_{1},\ldots,\rv_{\ell}\right)
    & \equiv \quantmean{
      \hat{\psi}_{g\nu_{1}}^{\dag}\left(\rv_{1}\right) \ldots
      \hat{\psi}_{g\nu_{\ell}}^{\dag}\left(\rv_{\ell}\right)
      \hat{\psi}_{g\tau_{\ell}}\left(\rv_{\ell}\right) \ldots
      \hat{\psi}_{g\tau_{1}}\left(\rv_{1}\right) }  \, \textrm{,}\\
    P_{\ell}^{\nu_{1}\tau_{1},\ldots,\nu_{\ell-1}\tau_{\ell-1};\nu_{\ell}\eta_{\ell}}\left(\rv_{1},\ldots,\rv_{\ell-1};\rv_{\ell}\right)
    & \equiv \quantmean{ \hat{\psi}_{g\nu_{1}}^{\dag}\left(\rv_{1}\right)
      \ldots
      \hat{\psi}_{g\nu_{\ell-1}}^{\dag}\left(\rv_{\ell-1}\right)
      \hat{\psi}_{g\nu_{\ell}}^{\dag}\left(\rv_{\ell}\right)
      \hat{\psi}_{e\eta_{\ell}}\left(\rv_{\ell}\right)
      \hat{\psi}_{g\tau_{\ell-1}}\left(\rv_{\ell-1}\right) \ldots
      \hat{\psi}_{g\tau_{1}}\left(\rv_{1}\right) } \, \textrm{.}
    \label{eq:CorrFuncDef_general}
  \end{align}
  The quantum-mechanical ground-level many-body correlation function
  $\rho_{\ell}^{\nu_{1}\tau_{1},\ldots,\nu_{\ell}\tau_{\ell}}\left(\rv_{1},\ldots,\rv_{\ell}\right)$
  represents the spatial correlations in the absence of incident light.
  These correlation functions are constants of motion in the low light
  intensity limit, and zero magnetic field.  The presence of a
  constant magnetic field induces a phase-rotation owing to Zeeman
  splitting.  Specifically,
  \begin{equation}
    \label{eq:rho_eqm_Zeeman}
    \frac{d}{dt}
    \rho_{\ell}^{\nu_{1}\tau_{1},\ldots,\nu_{\ell}\tau_{\ell}}\left(\rv_{1},\ldots,\rv_{\ell}\right)
    = i\left[
      \sum_{k=1}^{\ell}
        \bar{\Delta}_{g\nu_k g\tau_k}
    \right]
    \rho_{\ell}^{\nu_{1}\tau_{1},\ldots,\nu_{\ell}\tau_{\ell}}\left(\rv_{1},\ldots,\rv_{\ell}\right)
    \, \textrm{.}
  \end{equation}

  To calculate the optical response of the system to the lowest order in
  the electric field amplitude, we follow the evolution of the $\ell$-atom correlation
  function
  $P_{\ell}^{\nu_{1}\tau_{1},\ldots,\nu_{\ell-1}\tau_{\ell-1};\nu_{\ell}\eta_{\ell}}\left(\rv_{1},\ldots,\rv_{\ell-1};\rv_{\ell}\right)$
  for the ground-level atoms and/or ground-level coherences at positions
  $\rv_1,\ldots,\rv_{\ell-1}$, given there is a polarization at position
  $\rv_\ell$.
  These correlation functions satisfy the equations of motion ($\ell=1,\ldots,N$)
  \begin{align}
    & \left\{ \frac{d}{dt} - i \left[ \bar{\Delta}_{g\nu_{\ell}e\eta_{\ell}}
        +\sum_{k=1}^{\ell-1}
		\bar{\Delta}_{g\nu_{k} g\tau_{k}}
		  \right] +\gamma \right\}
    P_{\ell}^{\nu_{1}\tau_{1},\ldots,\nu_{\ell-1}\tau_{\ell-1};\nu_{\ell}\eta_{\ell}}\left(\rv_{1},\ldots,\rv_{\ell-1};\rv_{\ell}\right) =
    \nonumber\\
    & i \frac{\xi}{\mathcal{D}}
    \rho_{\ell}^{\nu_{1}\tau_{1},\ldots,\nu_{\ell-1}\tau_{\ell-1},\nu_{\ell}\tau}\left(\rv_{1},\ldots,\rv_{\ell}\right)
    \mathcal{C}_{\tau,\eta_{\ell}}^{\left(\sigma\right)}
    \pol_{\sigma}^{\ast} \cdot \Dv_{F}^{+}\left(\rv_{\ell}\right)
    \nonumber\\
    & + i \xi \sum_{k=1}^{\ell-1}
    \mathcal{C}_{\tau,\eta_{\ell}}^{\left(\sigma\right)}
    \pol_{\sigma}^{\ast} \cdot
    \radKernel^{\prime}\left(\rv_{\ell}-\rv_{k}\right)
    \pol_{\varsigma}
    \mathcal{C}_{\tau_{k},\eta}^{\left(\varsigma\right)}
    \nonumber \\
    & \quad\quad\quad \times
    P_{\ell}^{\nu_{1}\tau_{1},\ldots,\nu_{k-1}\tau_{k-1},\nu_{\ell}\tau,\nu_{k+1}\tau_{k+1},\ldots,\nu_{\ell-1}\tau_{\ell-1};\nu_{k}\eta}\left(\rv_{1},\ldots,\rv_{k-1},\rv_{\ell},\rv_{k+1},\ldots,\rv_{\ell-1};\rv_{k}\right)
    \nonumber\\
    & +i\xi \int
    d^{3}r_{\ell+1}\,\mathcal{C}_{\tau,\eta_{\ell}}^{\left(\sigma\right)}
    \pol_{\sigma}^{\ast}\cdot
    \radKernel^{\prime}\left(\rv_{\ell}-\rv_{\ell+1}\right)
    \pol_{\varsigma}
    \mathcal{C}_{\nu,\eta}^{\left(\varsigma\right)}
    P_{\ell+1}^{\nu_{1}\tau_{1},\ldots,\nu_{\ell}\tau;\nu\eta}\left(\rv_{1},\ldots,\rv_{\ell};\rv_{\ell+1}\right)
    \,\textrm{,}
    \label{eq:corrFuncEqsOfM_general}
  \end{align}
  where repeated indices $\nu$, $\tau$, $\eta$, $\sigma$, and
  $\varsigma$ are summed over.

Our primary goal in this Appendix is to show that the hierarchy of equations
\eqref{eq:corrFuncEqsOfM_general} can be solved exactly (within the
statistical accuracy) by classical electrodynamics simulations for the case of
a single electronic ground level.  (Here we specifically consider the
$J=0\rightarrow J'=1$ transition.)  These classical simulations account for
light-mediated interactions and recurrent (or dependent) scattering between
the atoms to all orders.
For atoms with an isotropic susceptibility with the $J=0 \rightarrow J'=1$ transition, we define the correlation
functions
\begin{align}
  \label{eq:CorrFuncDef}
    \Pv_\ell(\rv_1,\ldots,\rv_{\ell-1}; \rv_\ell) &\equiv
    \Dc \pol_\sigma
    P_\ell^{00,\ldots,00;0\sigma}(\rv_1,\ldots,\rv_N) \, \textrm{,}
    \\
    \rho_\ell(\rv_1,\ldots,\rv_\ell) &\equiv
    \rho_\ell^{00,\ldots,00}(\rv_1,\ldots,\rv_\ell) \, \textrm{,}
    \label{eq:DensityCorrFuncDef}
  \end{align}
  where the repeated index $\sigma$ is summed over.
  In this simplified case, the many-body ground-level correlation functions are
  constants of motion, and $\Pv_\ell$ satisfies the equations of
  motion
  \begin{align}
    \label{eq:corrFuncEqsOfM}
    \dot{\Pv}_\ell&(\rv_1,\ldots,\rv_{\ell-1}; \rv_\ell) = (i\bar{\Delta}  -
    \gamma)\Pv_{\ell} (\rv_1,\ldots,\rv_{\ell-1}; \rv_\ell) + i\xi
    \rho_\ell(\rv_1,\ldots,\rv_\ell)\Dv^+_F(\rv_\ell) \nonumber\\
    &  +  i\xi\sum_{k=1}^{\ell-1} \radKernel'(\rv_\ell - \rv_k)
    \Pv_\ell(\rv_1,\ldots,\rv_{k-1},\rv_{\ell},\rv_{k+1},\ldots,\rv_{\ell-1};\rv_k)
    + i\xi \int d^3\rv_{\ell+1} \radKernel'(\rv_\ell - \rv_{\ell+1})
    \Pv_{\ell+1}(\rv_1,\ldots,\rv_\ell;\rv_{\ell+1}) \, .
  \end{align}
Here $\Pv_\ell(\rv_1,\ldots,\rv_{\ell-1}; \rv_\ell)$ is a correlation function for
the ground-state atomic positions at $\rv_1,\ldots,\rv_{\ell-1}$, given that there
is a polarization at $\rv_\ell$.  The corresponding hierarchy for equations of motion governing two-level
atoms can be found from Eq.~\eqref{eq:corrFuncEqsOfM} by making the replacement
$\Pv_{\ell}  \rightarrow \pol P_{\ell} $.

In the limit of low light intensity $\rho_\ell$ are not affected by
the excitations and are constants of the motion.
The ground-state correlations are related to the many-body wave function of the atoms by
\beq
\rho_\ell (\rv_1,\ldots,\rv_\ell) = {N!\over (N-l)!} \int d^3 \rv_{\ell+1}\ldots d^3\rv_N \,
|\Psi(\rv_1,\ldots,\rv_N)|^2
\,.
\label{eq:rholwave}
\eeq
For a classical (initially uncorrelated ensemble) or, e.g., a Bose-Einstein
condensate, we simply have
\beq
\rho_\ell (\rv_1,\ldots,\rv_\ell) = {\cal N}\rho_1 (\rv_1)\ldots \rho_1
(\rv_\ell)\,,
\label{eq:rho1factorised}
\eeq
where ${\cal N}$ is a normalisation factor.

We introduce an ensemble of classical dipoles at fixed positions $\stochxv_j$ for $j=1,\ldots, N$. The individual atomic dipole amplitudes $\spvec{\Pc}^{(j)}$ then interact via the scattered electromagnetic field and evolve according to Eq.~\eqref{eq:Pol_single_atom_J0_J1},
\beq
  {d\over dt}  \spvec{\Pc}^{(j)} = (i\bar{\Delta}-\gamma) \spvec{\Pc}^{(j)}
  + i{\xi\over \Dc}\,  {\bf D}^+_F(\stochxv_j)  +i\xi \sum_{l\neq j}
  {\sf G}'(\stochxv_j-\stochxv_l) \spvec{\Pc}^{(l)} \, .
  \label{singlerealization_app}
\eeq
By means of treating the atomic positions $\{\stochxv_1,\ldots,\stochxv_N\}$ as random variables that
satisfy an appropriately chosen joint probability distribution $P(\rv_1,\ldots,\rv_N)$, we can construct an ensemble-averaged solution  that reproduces dynamics equivalent to the hierarchy of equations of motion~\eqref{eq:corrFuncEqsOfM}.
In practise, we solve the coupled equations
for the light and atomic dipoles for each stochastic realization of a fixed set of positions and ensemble average the quantities of interest over many such realizations.

We take the joint probability distribution of the positions of the atomic dipoles to be the absolute square of the quantum many-body wave function of the ground-state atoms
\begin{equation}
P(\rv_1,\ldots,\rv_N) = |\Psi(\rv_1,\ldots,\rv_N)|^2\,.
\label{PDT}
\end{equation}
With this choice, we may now introduce classical ground-state position correlation functions $\tilde\rho_\ell(\rv_1,\ldots,\rv_\ell) $ for the atoms that coincide with the normally-ordered quantum-mechanical position correlation functions $\rho_\ell (\rv_1,\ldots,\rv_\ell)$. Specifically,
\begin{align}
 \tilde\rho_\ell&(\rv_1,\ldots,\rv_\ell) \equiv \big\langle\sum_{j_1,\ldots,j_\ell}\!\!\!{}'
 \,\delta(\rv_1 -
    \stochxv_{j_1}) \ldots \delta(\rv_\ell -
      \stochxv_{j_\ell}) \big\rangle\, ,
    \label{eq:DensityCorrFun_local}\\
    &=  \sum_{j_1,\ldots,j_\ell}\!\!\!{}'\int d^3\stochxv_{1}\ldots d^3 \stochxv_{N}\,
P(\stochxv_{1},\ldots,\stochxv_{N})\delta(\rv_1- \stochxv_{j_1})\ldots\delta(\rv_\ell-\stochxv_{j_\ell})\nonumber\\
&= {N!\over (N-\ell)!} \int d^3 \rv_{\ell+1} \ldots d^3\rv_N\, P(\rv_1,\ldots,\rv_N)\,.
\end{align}
The prime in the summation indicates that those terms in which any of
$j_1,\ldots,j_\ell$ refer to the same atom are excluded. In the last line we used
the fact that the primed summations over the $\ell$-tuples $j_1,\ldots, j_\ell$
include all $N!/(N-\ell)!$ permutations of $\ell$ indices from the set of $N$
atomic indices $\{j=1,\ldots, N\}$. We can therefore  in principle simulate the
quantum-mechanical position correlations $\rho_\ell (\rv_1,\ldots,\rv_\ell)$ by
stochastic sampling of the positions of classical dipoles in the hierarchy of
equations \eqref{eq:corrFuncEqsOfM}. In order to show that the hierarchy can be
solved by integrating Eq.~\eqref{singlerealization_app} for each individual
stochastic realization and subsequently ensemble-averaging the results, we proceed
as follows. We consider an ensemble of atoms at fixed positions $\stochxv_j$ for
$j=1,\ldots, N$, multiply the terms in Eq.~\eqref{singlerealization_app} by
products of delta functions, and sum over the atomic positions to obtain
  \begin{align}
    \left(\frac{d}{dt}+\gamma-i\bar{\Delta}\right)&
       \Dc{\sum_{j_1,\ldots,j_\ell}}\!\!\!{}'
    \delta(\rv_1 - \stochxv_{j_1}) \ldots \delta(\rv_\ell - \stochxv_{j_\ell})
	\spvec{\Pc}^{(j_\ell)}
      = i{\xi}\,
       {\sum_{j_1,\ldots,j_\ell}}\!\!\!{}' \delta(\rv_1 -
    \stochxv_{j_1}) \ldots \delta(\rv_\ell -
      \stochxv_{j_\ell})
      \Dv^+_F(\rv_{\ell})\nonumber\\
    &+  i  \xi
    {\sum_{j_1,\ldots,j_\ell}}\!\!\!{}' \,\sum_{j_m\ne j_\ell}
   	\delta(\rv_1 - \stochxv_{j_1}) \ldots \delta(\rv_\ell -
	\stochxv_{j_\ell})\Dc\radKernel'(\rv_{\ell}-\stochxv_{j_m})\spvec{\Pc}^{(j_m)} \, .
	\label{eq:corr_func_eqM_deriv_1}
  \end{align}
There are two types of terms contained in the summation over $j_m$ in the last
line of Eq.~\eqref{eq:corr_func_eqM_deriv_1}. Either $j_m$ corresponds to an index of one of the atoms
in the $\ell$-tuples $(j_1,\ldots,j_\ell)$ that do not reside at $\rv_\ell$ (i.e.,
$j_m\in\{j_1,\ldots,j_{\ell-1}\}$), or it  does not. Accounting for each type of term separately and taking the expectation value, we find that the last line of Eq.~\eqref{eq:corr_func_eqM_deriv_1} becomes
  \begin{align}
    i  \xi &
    \big\langle {\sum_{j_1,\ldots,j_\ell}}\!\!\!{}' \,\sum_{j_m\ne j_\ell}
    \delta(\rv_1 - \stochxv_{j_1}) \ldots \delta(\rv_\ell -
	\stochxv_{j_\ell})\Dc\radKernel'(\rv_{\ell}-\stochxv_{j_m})\spvec{\Pc}^{(j_m)} \big\rangle \nonumber\\
  & = i\xi\sum_{k=1}^{\ell-1} \radKernel'(\rv_\ell - \rv_k)
    \tilde\Pv_\ell(\rv_1,\ldots,\rv_{k-1},\rv_{\ell},\rv_{k+1},\ldots,\rv_{\ell-1};\rv_k) \nonumber\\
  &  + i\xi \int d^3\rv_{\ell+1} \radKernel'(\rv_\ell - \rv_{\ell+1})
    \tilde\Pv_{\ell+1}(\rv_1,\ldots,\rv_\ell;\rv_{\ell+1}) \, ,
    \label{eq:eq:corr_func_eqM_deriv_2}
  \end{align}
where we have defined the classical correlation function in an analogy to Eq.~\eqref{eq:CorrFuncDef} as
\beq
    \label{eq:CorrFunc_StochRel_def}
    \tilde\Pv_\ell(\rv_1,\ldots,\rv_{\ell-1}; \rv_\ell) \equiv  \big\langle
	\Dc{\sum_{j_1,\ldots,j_\ell}}\!\!\!{}'
    \delta(\rv_1 - \stochxv_{j_1}) \ldots \delta(\rv_\ell - \stochxv_{j_\ell})
	\spvec{\Pc}^{(j_\ell)} \big\rangle\,.
\eeq
The second line of Eq.~\eqref{eq:eq:corr_func_eqM_deriv_2}
accounts for the field scattered from atoms at positions
$\rv_1,\ldots,\rv_{\ell-1}$ to the atom at $\rv_\ell$, and the third
line accounts for scattering from other atoms ($\ell+1,\ldots,N$) in the system to
the atom at position $\rv_\ell$.
Taking the average of all terms of Eq.~\eqref{eq:corr_func_eqM_deriv_1}, we obtain the same hierarchy of equations for the classical correlation functions $\tilde\Pv_\ell$ and
$\tilde\rho_\ell$, as for the quantum-mechanical correlations  $\Pv_\ell$ and $\rho_\ell$ in  Eq.~\eqref{eq:corrFuncEqsOfM}.

  The averages involved in the calculations
  can be evaluated by Monte-Carlo simulations.
  One takes many realizations of $N$ atomic positions
  $\stochxv_1,\ldots,\stochxv_N$ from the system's initial $N$-body
  density correlation function.
  Then, for each realization of random variables, we take
  the atoms to be localized at the sampled positions so that the
  correlation functions $\rho_\ell$ and $\Pv_\ell$ are described by
  Eq.~\eqref{eq:CorrFuncDef}.
  Then, one solves the dynamics of $\spvec{\Pc}^{(j)}$, which depend on
  the atomic positions.
  The correlation functions then emerge as their average over many
  stochastic realizations.
\end{widetext}

\section{Classical electrodynamics in the multilevel low light intensity case}
\label{app:multilevel}

Our general approach in this paper addresses spatial correlations in
the optical response by fixing the atomic positions in each stochastic
realization according to some given probability distribution and then
ensemble-averaging over many such realizations.  In the case of atoms
with multiple electronic ground levels, however, each spatial
configuration for the multilevel case still includes the internal level
dynamics. In
Sec.~\ref{sec:lowintensitymultilevel} we presented a formalism how to
incorporate the internal level dynamics in classical electrodynamics
simulations. The analysis of the internal level many-body correlations in
Sec.~\ref{sec:incorp-two-body} provides quantum corrections to the multilevel
dynamics in Eq.~\eqref{eq:multi_atom_2_body_eqm}. We will now show that
Eq.~\eqref{eq:multi_atom_2_body_eqm} involves two types of dynamical
processes: (i) classical scattering processes (for which $\nu=\tau$ and
$\mu=\beta$) that are already incorporated in the classical electrodynamics
simulation model of Sec.~\ref{sec:lowintensitymultilevel}; and (ii) virtual
quantum processes that go beyond the classical coupled dipole model.

Consider the low light intensity classical multilevel electrodynamics model of Eq.~\eqref{eq:Pcv_multi_level_dynamics},
\begin{align}
  {d\over dt} &\Pc_{\mu\eta}^{(j)} = (i\bar{\Delta}_{g\mu e\eta}-\gamma)\Pc_{\mu\eta}^{(j)}
  + i\frac{\xi}{\Dc}\,  \mathcal{C}_{\mu,\eta}^{(\sigma)} \pol_{\sigma}^\ast\cdot
  \Dv^+_F(\stochxv_j) \nonumber\\
  & +i\xi \sum_{l\neq j} {\cal C}_{\mu,\eta}^{(\sigma)}\pol_{\sigma}^\ast\cdot
  \radKernel'(\stochxv_j-\stochxv_l)  \pol_{\varsigma}\mathcal{C}_{\epsilon,\zeta}^{(\varsigma)}
 f_\epsilon^{(l)}\Pc_{\epsilon\zeta}^{(l)}\, \textrm{.}
  \label{eq:Pcv_multi_level_dynamics2}
\end{align}
where we have used the factorization of Eq.~\eqref{eq:multi_level_P2_simple},
\beq
\Pc_{\nu\tau;\epsilon\zeta}^{(j;l)} \simeq
  \delta_{\nu\tau}f_\nu^{(j)}f_\epsilon^{(l)}\Pc_{\epsilon\zeta}^{(l)}\,.
\label{appfacto}
\eeq
The formula \eqref{eq:Pcv_multi_level_dynamics2} provides the classical electrodynamics simulation model for time-dependent multilevel systems. Next we replace $j$ by $j_2$ and $l$ by $j_3$ in Eq.~\eqref{eq:Pcv_multi_level_dynamics2} and multiply the both sides of Eq.~\eqref{eq:Pcv_multi_level_dynamics2} by $f_\nu^{(j_1)}f_\mu^{(j_2)}$, indicating that the atom $j_1$ is in the level $\nu$ and $j_2$ in $\mu$. In the last term we have to deal separately with the terms $j_1=j_3$ and $j_1\neq j_3$. Noting that the atom $j_1$ occupies $\nu$, we obtain
\begin{widetext}
\begin{align}
  & \left[ \frac{d}{dt}-i\bar{\Delta}_{g\mu e\eta}  + \gamma\right]
  \Pc_{\nu\nu;\mu\eta}^{\left(j_{1};j_{2}\right)} = i\frac{\xi}{\mathcal{D}}
  f_{\nu}^{(j_1)}f_{\mu}^{(j_2)}
  \mathcal{C}_{\mu,\eta}^{\left(\sigma\right)}
  \pol_{\sigma}^{\ast} \cdot \Dv_{F}^{+}\left(\stochxv_{j_{2}}\right)  \nonumber\\
  & +i\xi
  \mathcal{C}_{\mu,\eta}^{\left(\sigma\right)}
  \pol_{\sigma}^{\ast} \cdot
  \radKernel^{\prime}\left(\stochxv_{j_{2}}-\stochxv_{j_{1}}\right)
  \pol_{\varsigma}
  \mathcal{C}_{\nu,\zeta}^{\left(\varsigma\right)}
  \Pc_{\mu\mu;\nu\zeta}^{\left(j_{2};j_{1}\right)} + i\xi \sum_{j_{3} \notin \{ j_{1},j_{2}\}}
  \mathcal{C}_{\mu,\eta}^{\left(\sigma\right)}
  \pol_{\sigma}^{\ast} \cdot
  \radKernel^{\prime}\left(\stochxv_{j_{2}}-\stochxv_{j_{3}}\right)
  \pol_{\varsigma}
  \mathcal{C}_{\epsilon,\zeta}^{\left(\varsigma\right)}
  \Pc_{\nu\nu,\mu\mu;\epsilon\zeta}^{\left(j_{1},j_{2};j_{3}\right)}\,,
  \label{eq:multi_atom_2_body_eqm_app}
\end{align}
\end{widetext}
where we have used the many-atom extension of Eq.~\eqref{appfacto}
\beq
\Pc_{\nu_1\nu_1,\ldots,\nu_{\ell-1}\nu_{\ell-1};\nu_\ell\zeta}^{(j_1,\ldots,j_{\ell-1};j_\ell)} = f_{\nu_1}^{(j_1)}\ldots f_{\nu_\ell}^{(j_\ell)}\Pc_{\nu_\ell\zeta}^{(j_\ell)}\,.
\label{newcor_app}
\eeq

Equation \eqref{eq:multi_atom_2_body_eqm_app} represents the case (i) of Eq.~\eqref{eq:multi_atom_2_body_eqm} and we can directly construct it from Eq.~\eqref{eq:Pcv_multi_level_dynamics2} by assuming the structure \eqref{newcor_app} for the correlation functions. In Eq.~\eqref{eq:multi_atom_2_body_eqm_app} the two-atom internal level correlation function $\Pc_{\nu\nu;\mu\eta}^{\left(j_{1};j_{2}\right)}$ is coupled to the three-atom internal level correlation function $\Pc_{\nu\nu,\mu\mu;\epsilon\zeta}^{\left(j_{1},j_{2};j_{3}\right)}$. Similarly, we can show that the three-atom internal level correlation function is coupled to the four-atom internal level correlation function. The process eventually yields a hierarchy of equations for the correlation functions.

Consider only the case (i) of Eq.~\eqref{eq:multi_atom_2_body_eqm} ($\nu=\tau$ and $\mu=\beta$), that is given by Eq.~\eqref{eq:multi_atom_2_body_eqm_app}, and the corresponding hierarchy of equations with only the analogous diagonal terms $\Pc_{\nu_1\nu_1,\ldots,\nu_{\ell-1}\nu_{\ell-1};\nu_\ell\zeta}^{(j_1,\ldots,j_{\ell-1};j_\ell)}$ included. In this case we can similarly extend our argument to show that Eqs.~\eqref{newcor_app} and~\eqref{eq:Pcv_multi_level_dynamics2} provide the full solution to the entire hierarchy of equations for internal level correlation functions. Therefore, the case (i) and all the correlation functions $\Pc_{\nu_1\nu_1,\ldots,\nu_{\ell-1}\nu_{\ell-1};\nu_\ell\zeta}^{(j_1,\ldots,j_{\ell-1};j_\ell)}$ can be solved by the classical electrodynamics simulations of Eq.~\eqref{eq:Pcv_multi_level_dynamics2} and the ansatz~\eqref{newcor_app}.
Note that, although the correlation function for the internal levels [Eq.~\ref{newcor_app}]  factorizes, the atomic system still becomes correlated by light due to induced spatial correlations. The case (ii) of Eq.~\eqref{eq:multi_atom_2_body_eqm} that does not necessarily have $\nu=\tau$ and $\mu=\beta$ is notably more complicated. The light-induced coupling in that case represents virtual quantum processes that are beyond the classical electrodynamics simulation method.

\end{document}